\documentclass[10pt]{article}

\usepackage[utf8]{inputenc} % allow utf-8 input
\usepackage[T1]{fontenc}    % use 8-bit T1 fonts
\usepackage{amsfonts}       % blackboard math symbols
\usepackage{microtype}
\usepackage{subfigure}
\usepackage{booktabs} % for professional tables
\usepackage{url}
\usepackage{times}
\usepackage{amsmath}
\usepackage{amsfonts}       % blackboard math symbols
\usepackage{latexsym}
\usepackage{array}
\usepackage{adjustbox}
\usepackage{multirow}
\usepackage{hyperref}
\usepackage{longtable}
\usepackage{soul}
\usepackage{enumitem}
\usepackage{xspace}
\usepackage{wrapfig}
\usepackage[font=small,labelfont=bf]{caption}
\usepackage{epstopdf}
\usepackage{courier}
\usepackage{placeins,makecell,authblk}
\usepackage{pifont,nccmath,bbm,titlesec}
\usepackage[square,numbers]{natbib}

\newcommand{\cmark}{\ding{51}}%
\newcommand{\xmark}{\ding{55}}%

\makeatletter
\newcommand*{\centerfloat}{%
  \parindent \z@
  \leftskip \z@ \@plus 1fil \@minus \textwidth
  \rightskip\leftskip
  \parfillskip \z@skip}
\makeatother

\usepackage[margin=1in]{geometry}

\makeatletter
\newcommand{\printfnsymbol}[1]{%
  \textsuperscript{\@fnsymbol{#1}}%
}
\makeatother

\newcommand{\RV}[1]{{#1}\xspace}

\title{SKM-TEA: A Dataset for Accelerated MRI Reconstruction with Dense Image Labels for Quantitative Clinical Evaluation}

\author{%
  Arjun D. Desai\footnote{\href{mailto:arjundd@stanford.edu}{\texttt{arjundd@stanford.edu}}}, Andrew M. Schmidt, Elka B. Rubin, Christopher M. Sandino,
  Marianne S. Black, Valentina Mazzoli, Kathryn J. Stevens, Robert Boutin, Christopher R\'e, Garry E. Gold, Brian A. Hargreaves, Akshay S. Chaudhari\\
  Stanford University\\
}
\date{\vspace{-5ex}}

\begin{document}

\maketitle

\begin{abstract}
    Magnetic resonance imaging (MRI) is a cornerstone of modern medical imaging. However, long image acquisition times, the need for qualitative expert analysis, and the lack of (and difficulty extracting) quantitative indicators that are sensitive to tissue health have curtailed widespread clinical and research studies. While recent machine learning methods for MRI reconstruction and analysis have shown promise for reducing this burden, these techniques are primarily validated with imperfect image quality metrics, which are discordant with clinically-relevant measures that ultimately hamper clinical deployment and clinician trust. To mitigate this challenge, we present the \textit{Stanford Knee MRI with Multi-Task Evaluation (SKM-TEA)} dataset, a collection of quantitative knee MRI (qMRI) scans that enables end-to-end, clinically-relevant evaluation of MRI reconstruction and analysis tools. This 1.6TB dataset consists of raw-data measurements of $\sim$25,000 slices (155 patients) of anonymized patient MRI scans, the corresponding scanner-generated DICOM images, manual segmentations of four tissues, and bounding box annotations for sixteen clinically relevant pathologies. We provide a framework for using qMRI parameter maps, along with image reconstructions and dense image labels, for measuring the quality of qMRI biomarker estimates extracted from MRI reconstruction, segmentation, and detection techniques. Finally, we use this framework to benchmark state-of-the-art baselines on this dataset. We hope our SKM-TEA dataset and code can enable a broad spectrum of research for modular image reconstruction and image analysis in a clinically informed manner. Dataset access, code, and benchmarks are available at \url{https://github.com/StanfordMIMI/skm-tea}.
\end{abstract}

\section{Introduction}
Magnetic resonance imaging (MRI) is a life-saving and sensitive tool for non-invasively diagnosing neurological, musculoskeletal, oncological, and other abnormalities \cite{vanBeek2018}. However, MRI data acquisition is inherently slow and can last up to an hour per patient, which can limit patient throughput in hospitals and can lead to increased patient wait times. Additionally, while MRI has enabled high-resolution anatomical imaging, identifying and quantifying pathology requires dense annotations (e.g. segmentations), which are cumbersome to curate and are prone to inter-reader variations. These limitations have stifled widespread and timely access to and analysis of MRI exams, leading to delayed or missed diagnoses while increasing the already burgeoning costs of healthcare.

Recently, machine learning (ML) has been utilized to create new algorithms for reconstructing high quality images while only sparsely sampling raw MRI data, which accelerates MRI scans and reduces overall scan time \cite{Hammernik2017}. Similarly, ML-based segmentation and detection tools have shown success in automating dense image labeling and in some cases, have even reached performance within the range of inter-reader variability of experts \cite{Bien2018}.

However, despite the high performance of new ML techniques for MRI reconstruction and analysis on existing benchmarks, few tools have successfully been deployed prospectively in clinics \cite{Chaudhari2020}. This translational barrier may be attributed to \textit{metric discordance} - the lack of agreement between true clinical utility of reconstructed MR images and annotations versus popular image quality analyzers (IQAs, e.g. peak signal-to-noise ratio [pSNR] and structural similarity [SSIM]), and pixel-wise or surface-based segmentation metrics (e.g. Dice, intersection over union [IOU], or average symmetric surface distance [ASSD]) \cite{desai2021international}. While these metrics provide a standardized and quantitative way of evaluating different techniques, the low sensitivity of these methods to clinically relevant features limit their utility in clinical decision-making. This issue is particularly prevalent in MRI reconstruction, where gold-standard evaluation requires time-consuming and expensive expert readings of images by radiologists. This has led to low concordance between radiologist assessments and ML metrics for determining image quality, and even competition-winning reconstruction models miss important pathologies that require surgical followup \cite{Knoll2020}. This problem is further exacerbated due to the lack of large-scale public datasets that include both the raw MRI data along \RV{with} multiple clinically-relevant observations for each image.

We seek to address these challenges with the \textit{Stanford Knee MRI with Multi-Task Evaluation (SKM-TEA)} dataset, a collection of raw MRI data, corresponding images, quantitative biomarkers, and dense tissue and pathology labels that together facilitate clinically relevant evaluation of MRI reconstruction and analysis methods. The main contributions of this work are as follows:
\begin{enumerate}[label={(\arabic*)}]
    \item \textbf{End-to-end MRI:} We curate a dataset of clinically-acquired quantitative MRI (qMRI) knee scans, images, and dense labels, which together can be used to generate clinically-meaningful, localized tissue-wise qMRI biomarker maps for end-to-end benchmarking of the MRI reconstruction and analysis pipeline.
    \item \textbf{Clinically-relevant evaluation:} We propose a standardized framework for evaluating model-generated outputs (reconstructions and dense labels) against these qMRI biomarker maps, which can serve as endpoints for quantitatively measures of tissue health.
    \item \textbf{Benchmarking:} We benchmark state-of-the-art MRI reconstruction and segmentation models using both traditional image and label metrics and the new clinically-relevant, quantitative MRI metrics enabled by this dataset.
\end{enumerate}

\begin{figure}[t]
  \centering
  \includegraphics[width=1\linewidth]{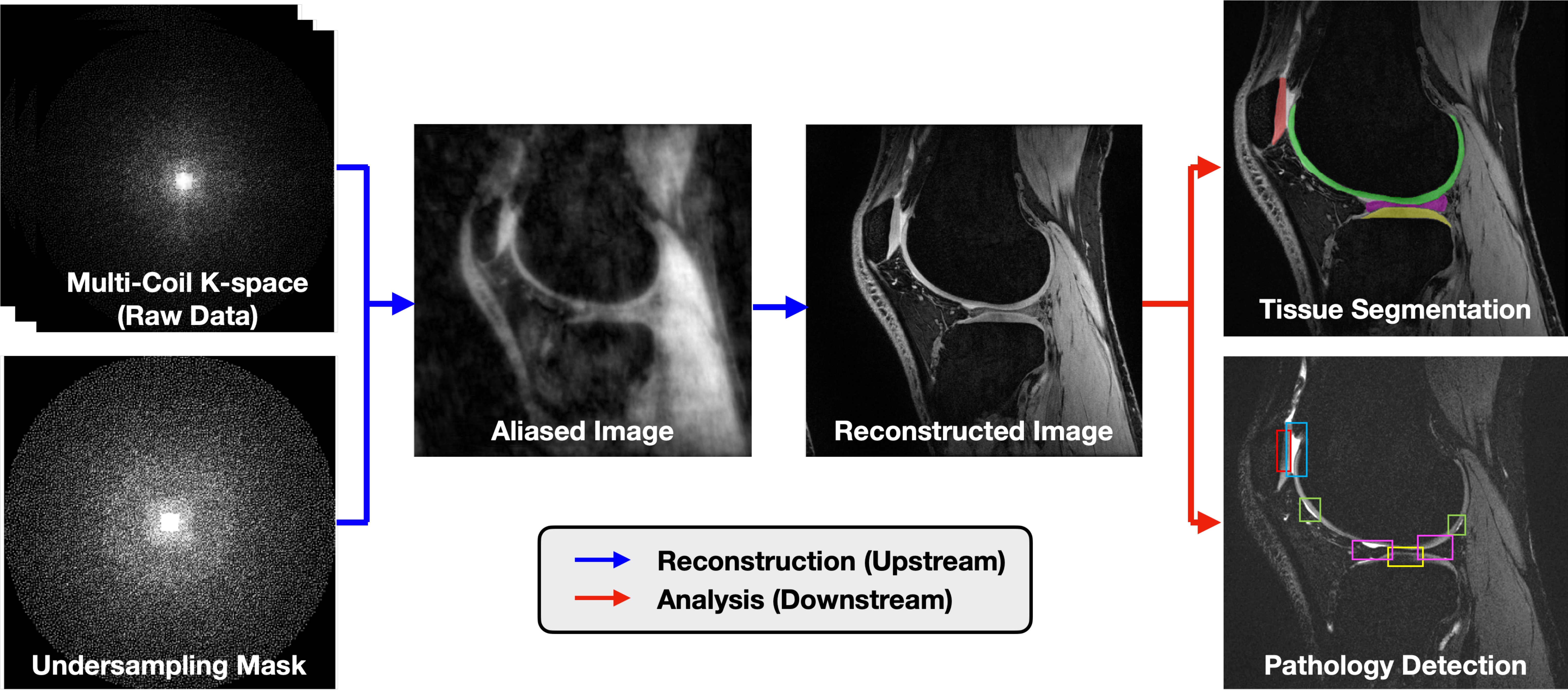}
  \caption{An overview of the end-to-end multi-coil MRI pipeline (and corresponding ML tasks). First, undersampled data acquired by multiple sensor coils is transformed into high quality images (i.e. reconstruction, blue arrow). Then, tissue regions of interest are localized (e.g. segmentation and detection) during image analysis. The SKM-TEA dataset curates raw data, ground-truth images, and dense annotations to enable all tasks. It also offers both a \textit{Raw Data Benchmarking Track}, which supports all these tasks, and the \textit{DICOM Benchmarking} track, which supports all image analysis tasks (red arrow).}
  \label{fig:overview}
\end{figure}

Dataset access, code, and evolving benchmarks are available at \url{https://github.com/StanfordMIMI/skm-tea}. Code is also distributed as a Python package: \texttt{pip install skm-tea}.

\section{Background}
Here, we provide a brief description of MRI reconstruction problem and introduce quantitative MRI.

\textbf{MRI reconstruction:} MRI scans require long acquisition durations because the raw-data that is acquired during imaging is the point-by-point Fourier transform of the desired image, termed as the \textit{k-space}. Sampling this k-space at the Nyquist sampling frequency is necessary to avoid image artifacts (i.e. aliasing). Accelerated MRI is a classical inverse problem consisting of subsampling the acquired raw k-space data below the Nyquist rate (that leads to artifactual images) and subsequently solving the ill-posed problem of recovering the original high-quality diagnostic image \cite{Hammernik2017}. 

\textbf{Quantitative MRI (qMRI):} Most MRI scans are fundamentally qualitative in nature, where clinical diagnoses are made based on \textit{relative} signal intensity differences between different image regions (i.e. are two adjacent tissues lighter/darker than normal?). Newer qMRI methods encode quantifiable physical and biochemical parameters directly into the images to ascribe a physical quantity per pixel, which enables cross-sectional and longitudinal studies of human health \cite{Chaudhari2019}. While qMRI methods are gaining popularity, their implementation is hampered due to extremely long scan times and the need to manually localize regions-of-interest (ROIs) in the image \cite{Peterfy2008, Miller2016, Petersen2009}. The availability of raw k-space data for qMRI studies will not only assist in defining new and improved local pixel-wise reconstruction accuracy metrics, but also encourage widespread adoption of rapid qMRI methods.

\begin{figure}[t]
  \centering
  \includegraphics[width=1.0\linewidth]{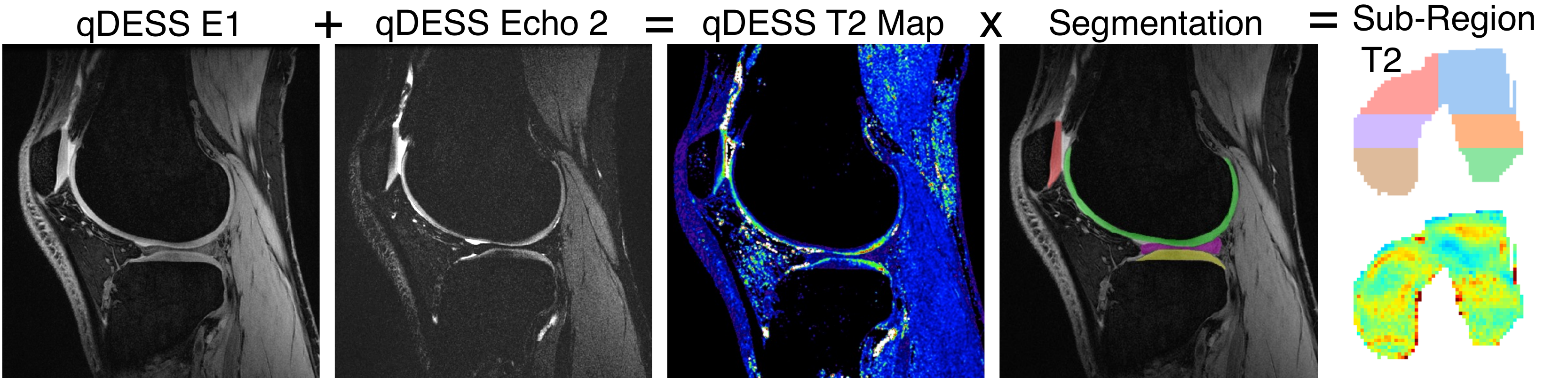}
  \caption{An overview of the qMRI parameter estimation pipeline. Reconstructed qDESS echo images are used to estimate quantitative $T_2$ parameters per pixel (i.e. $T_2$ maps). Tissue segmentations of articular cartilage and the meniscus are used to create tissue-wise $T_2$ maps. These segmentations are also automatically divided into sub-regions to get more localized regions of $T_2$ (example 3D femoral cartilage projected onto 2D). Regional $T_2$ estimates computed using ground truth data are compared to estimates produced using new methods for reconstruction or image segmentation.}
  \label{fig:qmri-eval-pipeline}
\end{figure}

The 3D quantitative double-echo steady-state (qDESS) MRI method allows acquiring rapid qualitative and quantitative knee imaging. qDESS acquires 2 sets of inherently-registered 3D images (termed echoes - E1 and E2) with varying image contrasts that can be used to compute the qMRI parameter of $T_2$ relaxation time (as shown in Fig. \ref{fig:qmri-eval-pipeline}), which is sensitive to collageneous tissue degeneration \cite{Sveinsson2017}. qDESS images have been used as a standalone 5-minute method for diagnostic knee MRI \cite{Chaudhari2018, Chaudhari2021} and can extract quantitative biomarkers for osteoarthritis in both knees \cite{Chaudhari2017, Eijgenraam2019, Kogan2017}.

\begin{table}[]
\centering
\caption{A summary of publicly available knee datasets, supported tasks, their size, and their inputs.}
\label{tbl:datasets-summary}
\begin{tabular}{lcccccc}
\toprule
\multirow{2}{*}{\textbf{Dataset}} & \multicolumn{4}{c}{\textbf{Tasks}} & \multirow{2}{*}{\textbf{Size (\textgreater{}20k slices)}} & \multirow{2}{*}{\textbf{qMRI}} \\
\cline{2-5}
& \textbf{Reconstruction} & \textbf{Classification} & \textbf{Segmentation} & \textbf{Detection} \\
\midrule
mridata \cite{epperson2013creation,ong2018mridata} & \cmark & \xmark & \xmark & \xmark & \xmark & \xmark \\
fastMRI(+) \cite{Knoll2020, zhao2021fastmri+} & \cmark & \cmark & \xmark & \cmark & \cmark & \xmark \\
MRNet \cite{Bien2018} & \xmark & \cmark & \xmark & \xmark & \cmark & \xmark \\
OAI \cite{Peterfy2008} & \xmark & \cmark & \cmark & \xmark & \cmark & \cmark \\
SKM-TEA & \cmark & \cmark & \cmark & \cmark & \cmark & \cmark \\
\bottomrule
\end{tabular}
\end{table}

\section{Related Work}
Previous datasets for MRI reconstruction, segmentation, and classification have been essential for enabling ML benchmarks for MRI research. However, these datasets are either limited in their size, application scope, or evaluation criteria. Table \ref{tbl:datasets-summary} summarizes the attributes of these datasets.

\textbf{Reconstruction:} mridata.org provides fully sampled raw 3D MRI k-space datasets for 19 healthy-subjects in an acquisition that took 40+ minutes each for scan \cite{epperson2013creation,ong2018mridata}. The small dataset size and homogeneous composition of healthy subjects makes it challenging to analyze how different reconstruction methods affect subtle pathology and clinical outcomes. Additionally, since this sequence is not commonly used in clinics, there exists a fundamental distribution shift between models built for this data acquisition and those that are clinically commonplace \cite{darestani2021}.
fastMRI is a large repository of k-space data acquired for $\approx$1600 2D knee and $\approx$7000 2D brain MRI scans of varying contrasts, with each imaging volume averaging 30-40 slices per acquisition \cite{Knoll2020dataset}. The extent of images made available through fastMRI and the global challenges has catalyzed ML research for MRI reconstruction \cite{Knoll2020, Muckley2021}. However, as described in the 2019 knee MRI reconstruction challenge, even for some of the challenge winning methods, conventional IQAs did not accurately convey true clinical imaging quality and also clearly obscured pathology \cite{Knoll2020}. This depicts a clear discordance between metrics used for ML evaluation and those for clinical image interpretation.

\textbf{Segmentation:} There are several datasets for dense tissue-level segmentations for MRI data from varying anatomies; however, none of these also include the raw k-space data that is required for MRI reconstruction \cite{Menze2015, tciaprostate}. Segmentation metrics are often also evaluated with quantitative volumetric or surface-based segmentation measures such as Dice and ASSD, instead of more clinically-relevant features. This often leads to a question regarding "how much performance is good enough?"  which cannot be answered without domain-specific clinical knowledge. Without image segmentation, extracting tissue-specific qMRI parameters of interest is also rendered impossible since most clinical use cases focus on specific tissues within an image, rather than the entire image. 

\textbf{Classification/Detection:} The MRNet and the Osteoarthritis Initiative (OAI) datasets provide labels for classification of knee MRI abnormalities \cite{Bien2018, Peterfy2008}. However, these classifications are only provided at a patient-level, and there are no abnormality bounding boxes to permit localization of the abnormality amongst multiple 3D imaging volumes. These datasets also do not provide the raw k-space, which hampers end-to-end evaluation of the impact of reconstruction methods on downstream clinical utility. Furthermore, the OAI datasets provides labels for the MRI Osteoarthritis Knee Score (MOAKS) on a subset of patients studied, which is similar, but not identical to routine clinical evaluation \cite{Hunter2011}. The fastMRI+ extension complements scans from the fastMRI reconstruction dataset with detection labels of clinical pathology \cite{zhao2021fastmri+}. While these labels are useful for localizing pathology in reconstructed images and for developing downstream detection models, evaluation still relies on standard IQA and detection metrics, which are discordant with clinically relevant endpoints.

\textbf{Overall Need:} Despite the large scale of data that has been made available, a limitation regarding optimal techniques for evaluating accelerated MRI reconstruction, especially for qMRI methods, still remains. Most datasets focus on a single specific evaluation task, which makes it challenging to simultaneously evaluate the benefit of acquisition or analysis methods. Segmentations to drive clinically-useful decisions exist for a handful of imaging datasets, but these have neither qMRI information nor raw k-space data. This motivates datasets that include raw k-space data for building image reconstruction methods with simultaneous evaluation metrics for clinically meaningful outcomes in a mutli-task manner to enable better cross-talk amongst the ML and medical imaging communities. 

\section{Dataset}

SKM-TEA consists of raw k-space MRI and image data collected from clinical knee qMRI scans, de-identified patient and imaging metadata, and corresponding dense tissue and pathology annotations for multi-task evaluation as shown in Fig. \ref{fig:overview}. In this section, we detail the data curation process for the raw data, images, and annotations as well as recommendations for distribution and usage. \RV{Additional details can be found in Appendix \ref{app:dataset}}.

\subsection{Data Collection}
\label{sec:dataset-collection}

\textbf{Collection overview:} 155 patients at Stanford Healthcare received a knee MRI with the qDESS sequence on one of two 3 Tesla (3T) GE MR750 scanners (GE, Waukesha, WI) with parameters shown in Table \ref{tbl:scan-parameters} (Appendix \ref{app:dataset-acq-params}). Both k-space data and scanner-generated DICOM images were collected, de-identified and stored securely. All data was collected with patient consent and Stanford University Institutional Review Board approval.

\textbf{Raw data (k-space):} All complex raw k-space data (with real and imaginary channels) were acquired in a multi-coil setting with 2x1 parallel imaging with elliptical sampling. Unsampled k-space data was subsequently synthesized using Autocalibrating Reconstruction for Cartesian imaging (ARC) with the GE Orchestra MATLAB SDK (v1.4) and was considered to be the fully-sampled k-space.

\textbf{DICOM images:} Standard reconstruction pipelines implemented by MRI vendors generate magnitude images that are distributed in the DICOM format \cite{parisot1995dicom,mildenberger2002introduction}. Since these methods may involve proprietary image and signal filtering, these images are not easy to reproduce. They are not interchangeable with images generated from the raw data and should not be used as a target for reconstruction tasks. However, scanner-generated DICOM images are useful for medical image analysis tasks (classification, segmentation, etc.) and are thus released with this dataset.

\textbf{$\mathbf{T_2}$ qMRI DICOM parameter maps}: Along with standard volumetric images, we also generated $T_2$ parameteric maps from the qDESS E1 and E2 images in the DICOM format using an analytical signal model described in \cite{Sveinsson2017}. Similar to the volumetric images, these underwent vendor-propriety post-processing. For this reason, these maps should only be used for visualization purposes. All analysis that uses $T_2$ maps should use the open-source implementation of this signal model in DOSMA (v0.1.0) \cite{desai2019dosma, arjun_desai_2019_3595808} to generate $T_2$ maps from the qDESS echoes.

\textbf{Tissue segmentations}: Manual segmentation for the following tissues were performed on all DICOM images: (1) patellar cartilage, (2) femoral cartilage, (3, 4) lateral and medial tibial cartilage, and (5, 6) lateral and medial meniscus.

\textbf{Localized pathology labels}: The DICOM images from all patients were reviewed alongside their radiology reports that described sixteen pathological categories across joint effusion and meniscal, ligament, and cartilage lesions. These reports were subsequently translated into 3D bounding boxes.

\subsection{Data Preparation}
\label{sec:dataset-prep}
\textbf{Hybridized k-space and SENSE reconstructions:} The 1D orthogonal inverse Fourier transform was applied to the fully-sampled k-space along the readout dimension to generate a hybridized k-space ($x \times k_y \times k_z$). As the readout direction is always fully-sampled, this operation alleviates memory constraints of modern computational accelerators. Sensitivity maps for each 2D ($k_y \times k_z$) slice were estimated using JSENSE (implemented in SigPy \cite{ong2019sigpy}) with a kernel-width of 6 and a 24$\times$24 center k-space auto-calibration region \cite{ying2007joint}. The fully-sampled k-space was then reconstructed using SENSE \cite{pruessmann1999sense}. We refer to the images reconstructed from the fully-sampled k-space and estimated sensitivity maps as \textit{SENSE reconstructions} to distinguish them from scanner-generated DICOM images.

Scanner-generated DICOM images undergo proprietary image filtering known as \textit{gradient warping corrections}, which leads to distortions between the SENSE-reconstructed images and the DICOM images (more details in Appendix \ref{app:grad-warp}) \cite{Doran2005}. Thus, all analysis or end-to-end inter-operation between SENSE reconstructions and tissue segmentations should use the gradient-warp-corrected segmentation. All image analysis starting only with the DICOM images should use the provided DICOM segmentations. 

\textbf{Dataset splits:} Of the 155 unique knee MRI scans, data from 36 patients who received additional arthroscopic surgical intervention were included in the test set. \RV{The remaining 119 scans were randomly split into 86 and 33 scans for training and validation, respectively}. Data from all splits is provided publicly. However, for benchmarking purposes, all models should only be trained using data in the training and validation splits. All metrics should be reported on the test split.

\textbf{Distribution:} Raw data, sensitivity maps, and SENSE reconstructions are distributed as pixel arrays in the HDF5 format. Image data and all segmentations are distributed in multiple formats (HDF5, DICOM, NIfTI) to facilitate interoperability between ML and clinical visualization workflows. Patient protected health information (PHI) metadata was anonymized from DICOM files, which were all subsequently manually inspected. Raw data, sensitivity maps, and scanner-generated and SENSE reconstructions were manually checked for PHI and quality. \RV{For additional details on distribution and maintenance, see Appendix \ref{app:dataset-distribution}}.

\section{Dataset Tracks}
\label{sec:dataset-tracks}

The SKM-TEA dataset enables two tracks with multi-task evaluations: (1) the Raw Data Benchmark and (2) the DICOM Image Benchmark. In this section, we discuss the available tasks in each track and best practices to ensure reproducibility for future work. Table \ref{tbl:data-per-track} summarizes the available data and the track(s) with which they are compatible.  When distributing or publishing benchmarks on this dataset, please report the specific dataset and \texttt{skm-tea} package versions used for all experiments (see Appendix \ref{app:dataset-usage} for usage details).

\subsection{Raw Data Benchmark Track}
The \textit{Raw Data Benchmark Track} relates to all tasks that leverage the raw k-space data and any artifacts generated from or related to the raw data. Data in this track enables tasks pertaining to MRI reconstruction, image segmentation, pathology detection, and qMRI parameter estimation.

\textbf{Reconstruction:} This task evaluates the quality of images reconstructed from undersampled (i.e. accelerated) MRI acquisitions. All model reconstructions are evaluated with respect to the complex-valued SENSE parallel-imaging reconstructions, which are state-of-the-art and clinically accepted, using fully-sampled raw data. Because all provided raw data, including that in the test set, is considered fully-sampled, k-space undersampling masks must be generated to simulate accelerated acquisitions. To ensure reproducibility among all test results, fixed undersampling masks are provided for each scan at acceleration factors of $R=4,6,8,10,12,16$. All undersampled inputs at test time should be generated using these fixed undersampling masks at the appropriate acceleration.

\textbf{Segmentation and Detection:} Tissue segmentation masks and localized pathology labels and bounding boxes can enable segmentation and detection tasks as well as localized image quality for evaluating reconstruction tasks in clinically pertinent regions. When using these labels in the context of this track, all image inputs correspond to ground-truth SENSE-reconstructed images or reconstructions generated directly from the raw data without additional post-processing. Gradient-warp corrected segmentations should be used as the segmentation masks in this track.

\textbf{Multi-Task MRI:} Given spatially-localized image labels are aligned with SENSE reconstruction targets, data in this track can enable multi-task learning setups that capture the end-to-end imaging workflow, from reconstruction to analysis (segmentation, detection, and qMRI generation).

\subsection{DICOM Benchmark Track}
The \textit{DICOM Benchmark Track} enables image analysis tasks using scanner-generated, magnitude DICOM images and corresponding tissue segmentations and detection labels. While data in this track is not compatible with reconstruction tasks, DICOM images have historically been the standard for downstream image analysis tasks like segmentation, detection, and classification \cite{Pace2015, Menze2015, desai2021international}.

\textbf{Segmentation and Detection}: Like in the previous benchmarking track, tissue and pathology labels can enable benchmarking of segmentation, detection, and other image analysis models. Segmentations without gradient-warp correction should be used for this track.

\section{Evaluation and Benchmarks}
\label{sec:eval-benchmarks}

\begin{table}
  \caption{Peak-signal-to-noise ratio (pSNR) and structural similarity (SSIM) [mean (standard deviation)] for SKM-TEA reconstruction baselines for echoes E1 and E2 accelerated (Acc.) at 6x and 8x.}
  \label{tbl:recon-iqa-perf}
  \centering
\begin{tabular}{llcccc}
\toprule
  & Metric & \multicolumn{2}{c}{pSNR (dB)} & \multicolumn{2}{c}{SSIM} \\
  \cmidrule(r){3-4} \cmidrule(r){5-6}
Acc.  & Model &           E1 &           E2 &              E1 &              E2 \\

%   & Echo &           E1 &           E2 &              E1 &              E2 \\
% Acc & Model &              &              &                 &                 \\

\midrule
\multirow{6}{*}{6x} & U-Net (E1/E2) &  31.5 (1.38) &  33.7 (1.02) &  0.77 (0.027) &             0.73 (0.032) \\
  & U-Net (E1+E2) &  31.1 (1.38) &  33.2 (1.05) &  0.77 (0.024) &  0.74 (0.030) \\
  & U-Net (E1$\oplus$E2) &  31.1 (1.63) &  33.5 (1.02) &  0.76 (0.026) &  0.73 (0.034) \\
  & Unrolled (E1/E2) &  \textbf{35.0 (1.08)} &  34.5 (1.09) &  0.83 (0.024) & 0.76 (0.031) \\
  & Unrolled (E1+E2) &  35.0 (1.07) &  \textbf{34.5 (1.09)} &  \textbf{0.84 (0.022)} &  \textbf{0.76 (0.030)} \\
  & Unrolled (E1$\oplus$E2) &  35.0 (1.08) &  34.2 (1.08) &  0.83 (0.023) &  0.76 (0.030) \\
\midrule
\multirow{6}{*}{8x} & U-Net (E1/E2) &  30.6 (1.55) &  32.9 (1.02) &  0.73 (0.030) &             0.67 (0.035)  \\
  & U-Net (E1+E2) &  30.8 (1.24) &  32.5 (1.00) &  0.72 (0.030) &  0.68 (0.035) \\
  & U-Net (E1$\oplus$E2) &  30.8 (1.23) &  32.7 (1.03) &  0.72 (0.029) &  0.68 (0.039) \\
  & Unrolled (E1/E2) &  33.8 (1.07) &  33.7 (1.06) &  0.79 (0.027) &  \textbf{0.73 (0.033)} \\
  & Unrolled (E1+E2) &  33.8 (1.07) &  33.6 (1.07) &  \textbf{0.80 (0.027)} &  0.71 (0.035) \\
  & Unrolled (E1$\oplus$E2) &  \textbf{33.9 (1.08)} &  \textbf{33.9 (1.07)} &  0.80 (0.027) &  0.73 (0.033) \\
\bottomrule
\end{tabular}
\end{table}

In this section, we introduce a standardized evaluation pipeline for using quantitative parameter map estimates and dense annotations to quantify clinically-relevant indicators as a new metric for performance of reconstruction and segmentation models. We use this metric, along with standard image quality and pixel-wise metrics, to benchmark models for a subset of tasks in each track.

\subsection{$\mathbf{T_2}$-based qMRI evaluation}
qMRI workflows to ascertain the biochemical status of tissues are highly sensitive to two key intermediate steps prior to retrieving clinically-relevant quantitative metrics: (1) image reconstructions must be high quality to allow for good parametric map estimation; and (2) regions of interest (ROIs) must be precisely segmented (or otherwise localized) to ensure per-pixel qMRI parameters are measured over the correct region. Due to their ability to measure disease endpoints and their sensitivity to these steps \cite{Chaudhari2020}, qMRI pipelines may be relevant targets for evaluating MRI reconstruction and image analysis methods.

We propose a qMRI-based evaluation framework that uses the $T_2$ parameter maps generated from the qDESS scans as a target for regional quantitative parameter analysis, as shown in Fig. \ref{fig:qmri-eval-pipeline}. In this pipeline, reconstructions of the two qDESS echoes are used to generate a $T_2$ parameter map. Tissue segmentations are subsequently applied to mask relevant ROIs in the parameter map to localize tissue-specific qMRI values. These ROIs can be further divided into physiologically-relevant sub-regions for each tissue. In the knee, the articular cartilage and meniscus, along with their sub-regions, have shown high predictive power for the early-onset of chronic disease \cite{Baum2013}. To extract the sub-regional parameters, tissue segmentations are automatically subdivided into these sub-regions, and $T_2$ parameters are averaged over these sub-regions (more sub-regional details in Appendix \ref{app:tissue-subregions}). To ensure reproducibility, all $T_2$ parameter map estimation and tissue sub-region division is done with the open-source implementation in DOSMA (v0.1.0). 

New reconstruction and image analysis methods can be used to generate image or label inputs into this pipeline. Differences in estimated $T_2$ values obtained from ground truth and predicted reconstructions and labels can be used to measure the accuracy of these estimates. \RV{Details on the pipeline used for each track can be found in Appendix \ref{app:qMRI-evaluation-pipeline}}.

\begin{table}
  \caption{Performance [mean (standard deviation)] of SKM-TEA reconstruction baselines with respect to absolute $T_2$ error (in milliseconds) for articular cartilage and the meniscus localized with ground truth segmentations. Typical cartilage $T_2$ values are 30-40ms, while meniscus $T_2$ values are 10-15ms).}
  \label{tbl:recon-t2-perf}
\centering
\begin{tabular}{llcccc}
\toprule
Acc. & Model &  Patellar Cartilage & Femoral Cartilage & Tibial Cartilage &     Meniscus \\
\midrule
\multirow{6}{*}{6x} & U-Net (E1/E2) &        2.19 (1.68) &       1.08 (0.94) &      1.61 (0.95) &  2.70 (1.35) \\
  & U-Net (E1+E2) &        2.83 (1.95) &       2.46 (1.88) &      1.46 (0.92) &  2.01 (1.42) \\
  & U-Net (E1$\oplus$E2) &        1.77 (1.50) &       1.11 (0.78) &      1.54 (1.03) &  1.81 (0.97) \\
  & Unrolled (E1/E2) &       \textbf{0.563 (0.23)} &      \textbf{0.765 (0.28)} &      \textbf{1.03 (0.42)} &  2.48 (0.79) \\
  & Unrolled (E1+E2) &       0.570 (0.23) &      0.836 (0.32) &      1.12 (0.42) &  2.52 (0.78) \\
  & Unrolled (E1$\oplus$E2) &        1.69 (1.36) &       2.01 (0.92) &      1.34 (0.55) &  \textbf{1.31 (0.82)} \\
\midrule
\multirow{6}{*}{8x} & U-Net (E1/E2) &        3.48 (1.74) &       2.71 (1.37) &      3.21 (1.24) &  3.76 (1.10) \\
  & U-Net (E1+E2) &        2.66 (2.06) &       3.04 (2.03) &      1.49 (1.16) &  \textbf{2.39 (1.31)} \\
  & U-Net (E1$\oplus$E2) &        1.29 (1.09) &       1.26 (0.91) &      2.09 (1.13) &  2.49 (1.80) \\
  & Unrolled (E1/E2) &       \textbf{0.721 (0.30)} &      \textbf{0.899 (0.34)} &      \textbf{1.26 (0.49)} &  2.78 (0.87) \\
  & Unrolled (E1+E2) &       0.971 (0.42) &      0.988 (0.39) &      1.30 (0.49) &  2.86 (0.88) \\
  & Unrolled (E1$\oplus$E2) &       0.588 (0.29) &      0.992 (0.43) &      1.33 (0.63) &  2.73 (0.89) \\
\bottomrule
\end{tabular}
\end{table}

% \footnote{It is important to note that qMRI analysis pipelines are decentralized and often differ in implementation. As a result, these workflows are often difficult to reproduce. DOSMA provides a reproducible framework for performing this analysis, but by no means is the only way of performing this analysis.}.
% \vspace{-3mm}
\subsection{Baselines}
We benchmark and summarize popular state-of-the-art models for MR reconstruction and segmentation on SKM-TEA below (more details on training setup in Appendix \ref{app:training-details}).

\textbf{Raw Data Track -- Reconstruction:} 2D U-Net \cite{ronneberger2015u} and unrolled \cite{diamond2017unrolled, sandino2020compressed} networks were trained to reconstruct 2D undersampled, complex-valued axial slices at 6x and 8x acceleration. As each scan consists of two 3D images (E1 and E2 echoes) for reconstruction, models were trained with the following configurations: (1) two separate models for echoes 1 and 2 (E1/E2); (2) a single model for both echoes, with each echo a unique training example (E1+E2); or (3) a single model for both echoes, with echoes 1 and 2 as multiple channels in a single example (E1$\oplus$E2).

% U-Net \& Unrolled, axial reconstruction, 2D poisson disc undersmapling, 6x and 8x. 3 configuration: E1/E2 separate, E1+E2 single model, or E1$\oplus$E2 multi-channel. trained on both v100 and titan rtx gpus but made sure not to exceed 16GB limit across all runs

\textbf{DICOM Track -- Segmentation:} 2D V-Net \cite{milletari2016v} and U-Net models were trained on DICOM images to segment patellar cartilage, femoral cartilage, tibial cartilage, and the meniscus. Separate models were trained on: (1) echo 1 only (E1), (2) echo 2 only (E2), (3) multi-channel echo1-echo2 (E1$\oplus$E2), and (4) the root-sum-of-squares (RSS) of the two echos.

% The model architecture was adopted from \cite{desai2019technical}.

\subsection{Metrics}
Standard image quality and segmentation metrics, in addition to the proposed $T_2$ evaluation framework, were used to evaluate reconstruction and segmentation models, respectively. Image reconstruction performance was measured using peak-signal-to-noise ratio (pSNR) and structural similarity (SSIM) on both qDESS echoes. Segmentation performance was measured with dice similarity coefficient (DSC), average symmetric surface distance in millimeters (ASSD, mm), volumetric overlap error (VOE), and coefficient of variation (CV).  Both $T_2$ error ($T_2^{pred} - T_2^{gt}$) and absolute $T_2$ error ($|T_2^{pred} - T_2^{gt}|$) (and their bias/variance) were used to evaluate estimated $T_2$ precision and accuracy.

\begin{figure}[t]
  \centering
  \includegraphics[width=1.0\linewidth]{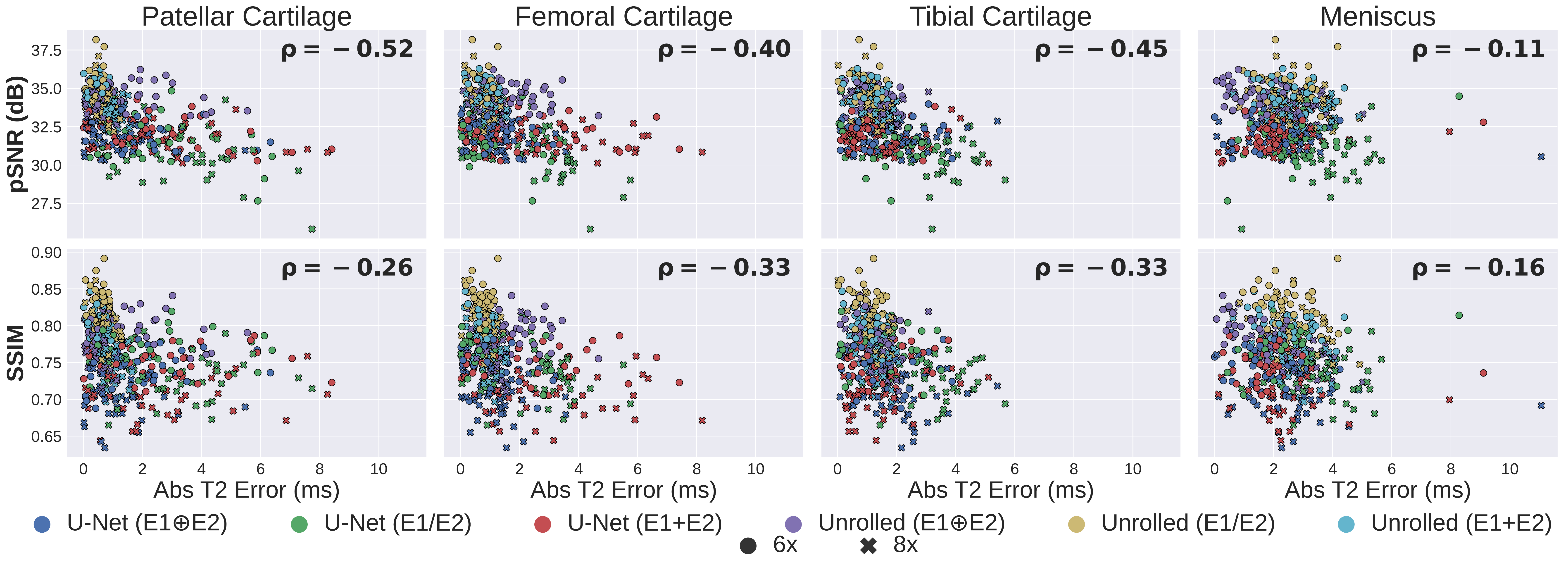}
  \caption{Performance among various Raw Data Track reconstruction models (colored markers) on image quality metrics versus absolute $T_2$ estimation error at both 6x (circle) and 8x (cross) acceleration. Top -- peak-signal-to-noise ratio (pSNR), bottom -- structural similarity (SSIM). Both pSNR and SSIM were very weakly correlated ($|\rho| \leq 0.52$) with $T_2$ estimation errors across all tissues.}
  \label{fig:corr-recon-t2}
\end{figure}

\subsection{Results and Analysis}
\textbf{Raw Data Track -- Reconstruction:} Among reconstruction models, unrolled models outperformed U-Net models in both pSNR and SSIM at both accelerations and across both echoes (Table \ref{tbl:recon-iqa-perf}). \RV{Unlike standard fully convolutional networks (e.g. U-Net), which rely on statistical imaging priors learned during training, unrolled networks can encode prior information about the image formation model with proximal update steps, which can improve image recovery. In addition,} among models of the same architecture, there was a negligible performance difference in how multi-echo inputs were input into the models. While standard IQA metrics indicated reasonably high performance, $T_2$ error performance of these models was more variable (Table \ref{tbl:recon-t2-perf}). U-Net (E1+E2) had the least bias for patellar and tibial cartilage $T_2$ estimates at 6x and for patellar cartilage at 8x. U-Net approaches had  higher variance (>1.0 ms) in these estimates compared to unrolled models. Models trained and evaluated on 6x-accelerated scans predominantly had lower variance, but higher bias in $T_2$ estimates. However, Unrolled (E1$\oplus$E2) had higher accuracy and lower variance in $T_2$ estimates of patellar cartilage at 8x-acceleration compared to 6x-acceleration. \RV{Given the higher signal in articular cartilage in E1 compared to E2, joint optimization of multi-channel (E1$\oplus$E2) inputs may result in implicit over-prioritization of reconstruction of the E1. This may lead to more variable per-pixel recovery for E2. As a result, per-pixel T2 maps, which are computed based on pixel-wise ratios between E2/E1, may have more variability and error. At higher acceleration factors, networks optimized with the l1-loss tend to blur the image (i.e. average the signal). Signal averaging may reduce the per-pixel variance and lead to more accurate region-wide signal, and thus T2, estimates}.  

\textbf{DICOM Track -- Segmentation:} Among all segmentation approaches, E1, $E1\oplus E2$, and RSS models had the highest and similar performance. E2 had consistently worse performance on both pixel-wise metrics and $T_2$ accuracy measures, likely due to the worse E2 echo image contrast. Additionally, low SNR of E2 compared to E1 may result in higher variability in segmentations, and thus more variability in $T_2$ accuracy. Other models had similar average performance and variance across DSC, VOE, and CV metrics. All approaches overestimated $T_2$ in patellar cartilage, femoral cartilage, and tibial cartilage, but underestimated $T_2$ in the meniscus. These methods also had higher variance (>0.6ms), which may indicate higher variability in the estimates despite low bias. V-Net models achieved higher performance compared to U-Net models among standard ML segmentation metrics (DSC -- patellar cartilage, ASSD -- all tissues), but had similar performance among $T_2$ error metrics. The discordance between relative differences for ML and qMRI metrics may suggest that standard ML metrics may not be a wholistic representation of relative performance differences between models. Thus, the use of only ML metrics for benchmark comparison may lead to inconsistencies between reported performance and practical utility. Results from U-Net benchmarks and additional metrics are detailed in Appendix \ref{app:additional-seg-baselines}.

\begin{figure}[t]
  \centering
  \includegraphics[width=1.0\linewidth]{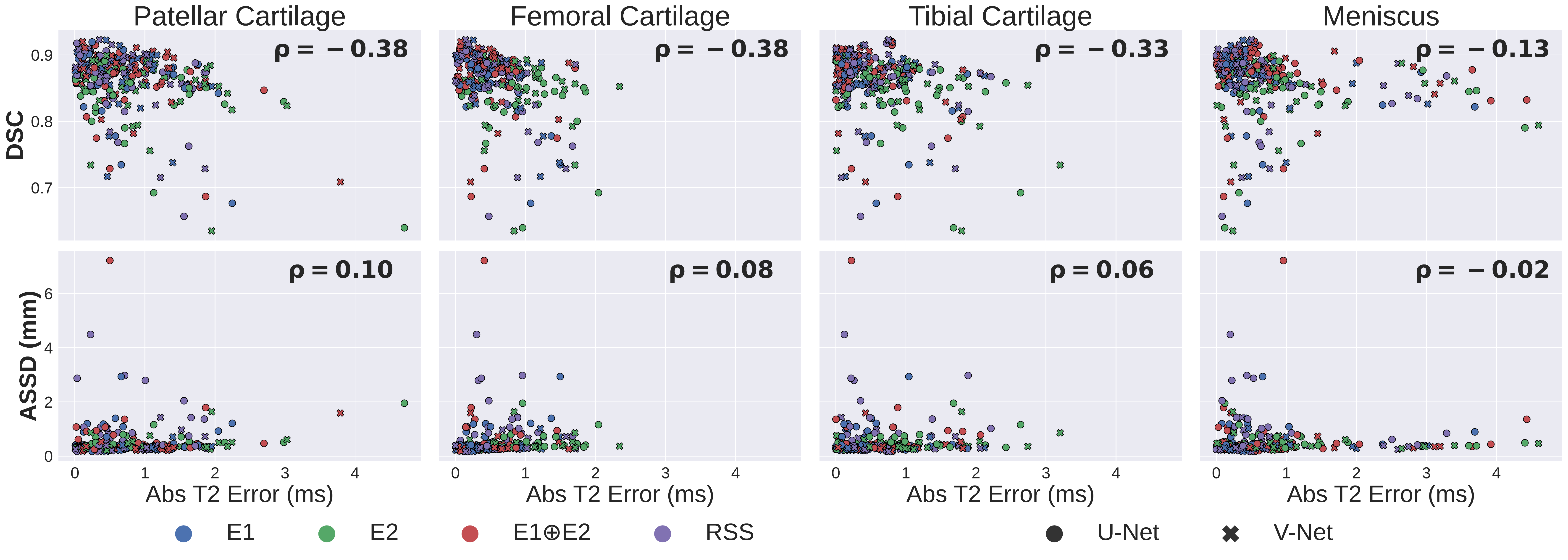}
  \caption{Performance among DICOM Track segmentation models (colored markers) on image quality metrics versus absolute $T_2$ estimation error. Both volumetric (top: dice score coefficient - DSC) and surface distance (bottom: average symmetric surface distance - ASSD) metrics were very weakly correlated ($|\rho| \leq 0.4$) with absolute errors in $T_2$ estimates.}
  \label{fig:corr-seg-t2}
\end{figure}

\textbf{Metric Concordance:} To understand concordance between standard ML metrics and $T_2$ accuracy, we quantified the Pearson correlation coefficient ($\rho$) between image quality and pixel-wise metrics with absolute $T_2$ estimation error across all models. Reconstruction metrics had very weak correlation with $T_2$ estimation error across all four tissues (both $|p| \leq 0.52$). A similar trend was observed for sub-regions of each of these tissues (Appendix \ref{app:t2-corr-subregions}). Both segmentation metrics (DSC and ASSD) were also very weakly correlated (both $|p| \leq 0.4$) with absolute $T_2$ error (Fig.\ref{fig:corr-seg-t2}). Thus, using clinically-relevant $T_2$ biomarkers as direct endpoints for quantifying performance can help mitigate the challenge of low concordance between standard ML metrics and accuracy of $T_2$ estimates. 

\begin{table}
  \caption{V-Net segmentation performance measured by standard ML pixel and surface segmentation metrics with absolute $T_2$ error. Models were trained with echo 1 only (E1), echo 2 only (E2), multi-channel echo 1 and echo 2 (E1$\oplus$E2), and the root-sum-of-squares (RSS) of both echoes.}
  \label{tbl:seg-perf}
  \centering
\begin{tabular}{llcccc}
\toprule
             Metric & Tissue &   E1 &    E2 & E1$\oplus$E2 &  RSS \\
\midrule
\multirow{4}{*}{DSC} & Patellar Cartilage &  0.88 (0.082) &   0.85 (0.11) &         \textbf{0.89 (0.084)} &  0.88 (0.088) \\
                  & Femoral Cartilage &  0.88 (0.035) &  0.86 (0.033) &         \textbf{0.88 (0.029)} &  0.88 (0.033) \\
                  & Tibial Cartilage &  0.86 (0.036) &  0.83 (0.048) &         0.86 (0.034) & \textbf{ 0.86 (0.034)} \\
                  & Meniscus &  0.85 (0.059) &  0.83 (0.052) &         \textbf{0.85 (0.056)} &  0.85 (0.060) \\
\midrule
\multirow{4}{*}{ASSD (mm)} & Patellar Cartilage &   \textbf{0.33 (0.28)} &   0.49 (0.63) &          0.36 (0.64) &   0.36 (0.54) \\
                  & Femoral Cartilage &   0.26 (0.11) &  0.29 (0.088) &         \textbf{0.25 (0.081)} &  0.25 (0.096) \\
                  & Tibial Cartilage &   0.33 (0.11) &   0.41 (0.17) &          0.32 (0.10) &  \textbf{0.32 (0.089)} \\
                  & Meniscus &   \textbf{0.49 (0.27)} &   0.54 (0.23) &          0.49 (0.29) &   0.54 (0.50) \\
\midrule
\multirow{4}{*}{Abs T2 Error (ms)} & Patellar Cartilage &   \textbf{0.64 (0.47)} &   1.02 (0.76) &          0.80 (0.71) &   0.66 (0.48) \\
                  & Femoral Cartilage &   0.53 (0.38) &   0.87 (0.51) &          \textbf{0.49 (0.35)} &   0.55 (0.40) \\
                  & Tibial Cartilage &   0.49 (0.51) &   0.87 (0.76) &          \textbf{0.47 (0.51)} &   0.52 (0.52) \\
                  & Meniscus &   \textbf{0.52 (0.63)} &    0.91 (1.0) &          0.71 (0.82) &   0.58 (0.66) \\
\bottomrule
\end{tabular}
\end{table}

\section{Limitations and Ethical Considerations}
\label{sec:limitations}
First, all data was acquired from patients who received a knee MRI scan at Stanford Hospital, which is a specific subgroup of the general population based on geographic, demographic, and health insurance constraints. Second, although data was collected from two different scanners, both scanners were from the same vendor, which may affect model performance on data acquired from different vendors. Given the heterogeneity of patient anatomy, and MRI acquisition and processing techniques, researchers training models on our dataset should account for potential distribution shifts and validate their models on relevant data sampled from their imaging setup prior to deployment. Third, although E2E-VarNet is a popular architecture for MRI reconstruction, it requires training data with a fixed number of coils for learning sensitivity maps \cite{sriram2020end}. However, scans in SKM-TEA represent real-world data where scans are acquired with different receiver coils with a heterogeneous number of coil elements due to variability in subject sizes. In future work, we will train networks on a subset of scans with the same number of coils. Additionally, there are other anatomical structures in the knee (e.g. bone, muscles), which are not densely annotated, but can help with extracting other clinically-relevant biomarkers. In future work, we will look to curate data from a larger subject pool and multi-vendor scanners and add annotations for these tissues.

\section{Conclusion}
In this work, we introduce SKM-TEA, a quantitative MRI knee dataset that enables clinically-relevant benchmarking of the end-to-end MRI workflow. First, we curate raw data, images, and dense tissue and pathology annotations from 155 clinical MRI scans. Second, we introduce two unique, but complementary, tracks for benchmarking MRI reconstruction, segmentation, and detection methods. Third, we propose an open-source evaluation framework that uses regional qMRI biomarker analysis as a direct endpoint for quantifying model performance. Finally, we train and evaluate state-of-the-art image reconstruction and analysis models on this dataset using the proposed evaluation framework. We find that existing evaluation metrics for both image and segmentation quality are discordant with qMRI biomarkers, particularly in physiologically relevant tissue subregions, which may warrant the need for the proposed evaluation framework for direct estimation of these biomarkers. 

By proposing a new dataset, two benchmarking tracks, and an evaluation framework using clinically-relevant qMRI biomarkers in a multi-task manner, we hope that SKM-TEA can enable a wide range of research in methods development and metric design across all stages of the MR imaging pipeline.

{
\small
\bibliographystyle{plain}
\bibliography{refs}

\begin{thebibliography}{10}

\bibitem{Baum2013}
T.~Baum, G.B. Joseph, D.C. Karampinos, P.M. Jungmann, T.M. Link, and J.S.
  Bauer.
\newblock Cartilage and meniscal t2 relaxation time as non-invasive biomarker
  for knee osteoarthritis and cartilage repair procedures.
\newblock {\em Osteoarthritis and Cartilage}, 21(10):1474--1484, October 2013.

\bibitem{Bien2018}
Nicholas Bien, Pranav Rajpurkar, Robyn~L. Ball, Jeremy Irvin, Allison Park,
  Erik Jones, Michael Bereket, Bhavik~N. Patel, Kristen~W. Yeom, Katie
  Shpanskaya, Safwan Halabi, Evan Zucker, Gary Fanton, Derek~F. Amanatullah,
  Christopher~F. Beaulieu, Geoffrey~M. Riley, Russell~J. Stewart, Francis~G.
  Blankenberg, David~B. Larson, Ricky~H. Jones, Curtis~P. Langlotz, Andrew~Y.
  Ng, and Matthew~P. Lungren.
\newblock Deep-learning-assisted diagnosis for knee magnetic resonance imaging:
  Development and retrospective validation of {MRNet}.
\newblock {\em {PLOS} Medicine}, 15(11):e1002699, November 2018.

\bibitem{Chaudhari2017}
Akshay~S. Chaudhari, Marianne~S. Black, Susanne Eijgenraam, Wolfgang Wirth,
  Susanne Maschek, Bragi Sveinsson, Felix Eckstein, Edwin~H.G. Oei, Garry~E.
  Gold, and Brian~A. Hargreaves.
\newblock Five-minute knee {MRI} for simultaneous morphometry and t2
  relaxometry of cartilage and meniscus and for semiquantitative radiological
  assessment using double-echo in steady-state at 3t.
\newblock {\em Journal of Magnetic Resonance Imaging}, 47(5):1328--1341,
  November 2017.

\bibitem{Chaudhari2021}
Akshay~S. Chaudhari, Murray~J. Grissom, Zhongnan Fang, Bragi Sveinsson,
  Jin~Hyung Lee, Garry~E. Gold, Brian~A. Hargreaves, and Kathryn~J. Stevens.
\newblock Diagnostic accuracy of quantitative multicontrast 5-minute knee {MRI}
  using prospective artificial intelligence image quality enhancement.
\newblock {\em American Journal of Roentgenology}, 216(6):1614--1625, June
  2021.

\bibitem{Chaudhari2019}
Akshay~S. Chaudhari, Feliks Kogan, Valentina Pedoia, Sharmila Majumdar,
  Garry~E. Gold, and Brian~A. Hargreaves.
\newblock Rapid knee {MRI} acquisition and analysis techniques for imaging
  osteoarthritis.
\newblock {\em Journal of Magnetic Resonance Imaging}, 52(5):1321--1339,
  November 2019.

\bibitem{Chaudhari2020}
Akshay~S. Chaudhari, Christopher~M. Sandino, Elizabeth~K. Cole, David~B.
  Larson, Garry~E. Gold, Shreyas~S. Vasanawala, Matthew~P. Lungren, Brian~A.
  Hargreaves, and Curtis~P. Langlotz.
\newblock Prospective deployment of deep learning in {MRI}: A framework for
  important considerations, challenges, and recommendations for best practices.
\newblock {\em Journal of Magnetic Resonance Imaging}, 54(2):357--371, August
  2020.

\bibitem{Chaudhari2018}
Akshay~S. Chaudhari, Kathryn~J. Stevens, Bragi Sveinsson, Jeff~P. Wood,
  Christopher~F. Beaulieu, Edwin~H.G. Oei, Jarrett~K. Rosenberg, Feliks Kogan,
  Marcus~T. Alley, Garry~E. Gold, and Brian~A. Hargreaves.
\newblock Combined 5-minute double-echo in steady-state with separated echoes
  and 2-minute proton-density-weighted 2d {FSE} sequence for comprehensive
  whole-joint knee {MRI} assessment.
\newblock {\em Journal of Magnetic Resonance Imaging}, 49(7):e183--e194,
  December 2018.

\bibitem{tciaprostate}
Peter Choyke, Baris Turkbey, Peter Pinto, Maria Merino, and Brad Wood.
\newblock Data from prostate-mri, 2016.

\bibitem{crowder2021characterizing}
Hollis~A Crowder, Valentina Mazzoli, Marianne~S Black, Lauren~E Watkins, Feliks
  Kogan, Brian~A Hargreaves, Marc~E Levenston, and Garry~E Gold.
\newblock Characterizing the transient response of knee cartilage to running:
  Decreases in cartilage t2 of female recreational runners.
\newblock {\em Journal of Orthopaedic Research{\textregistered}}, 2021.

\bibitem{darestani2021}
Mohammad~Zalbagi Darestani, Akshay~S Chaudhari, and Reinhard Heckel.
\newblock Measuring robustness in deep learning based compressive sensing.
\newblock In Marina Meila and Tong Zhang, editors, {\em Proceedings of the 38th
  International Conference on Machine Learning}, volume 139 of {\em Proceedings
  of Machine Learning Research}, pages 2433--2444. PMLR, 18--24 Jul 2021.

\bibitem{arjun_desai_2019_3595808}
Arjun Desai, Akshay Chaudhari, and Marco Barbieri.
\newblock ad12/dosma, December 2019.

\bibitem{desai2019dosma}
Arjun~D Desai, Marco Barbieri, Valentina Mazzoli, Elka Rubin, Marianne~S Black,
  Lauren~E Watkins, Garry~E Gold, Brian~A Hargreaves, and Akshay~S Chaudhari.
\newblock Dosma: A deep-learning, open-source framework for musculoskeletal mri
  analysis.
\newblock In {\em Proc. Intl. Soc. Mag. Reson. Med}, volume~27, 2019.

\bibitem{desai2021international}
Arjun~D Desai, Francesco Caliva, Claudia Iriondo, Aliasghar Mortazi, Sachin
  Jambawalikar, Ulas Bagci, Mathias Perslev, Christian Igel, Erik~B Dam, Sibaji
  Gaj, et~al.
\newblock The international workshop on osteoarthritis imaging knee mri
  segmentation challenge: a multi-institute evaluation and analysis framework
  on a standardized dataset.
\newblock {\em Radiology: Artificial Intelligence}, 3(3):e200078, 2021.

\bibitem{desai2019technical}
Arjun~D Desai, Garry~E Gold, Brian~A Hargreaves, and Akshay~S Chaudhari.
\newblock Technical considerations for semantic segmentation in mri using
  convolutional neural networks.
\newblock {\em arXiv preprint arXiv:1902.01977}, 2019.

\bibitem{diamond2017unrolled}
Steven Diamond, Vincent Sitzmann, Felix Heide, and Gordon Wetzstein.
\newblock Unrolled optimization with deep priors.
\newblock {\em arXiv preprint arXiv:1705.08041}, 2017.

\bibitem{Doran2005}
Simon~J Doran, Liz Charles-Edwards, Stefan~A Reinsberg, and Martin~O Leach.
\newblock A complete distortion correction for {MR} images: I. gradient warp
  correction.
\newblock {\em Physics in Medicine and Biology}, 50(7):1343--1361, March 2005.

\bibitem{Driban2014}
Jeffrey~B. Driban, Charles~B. Eaton, Grace~H. Lo, Robert~J. Ward, Bing Lu, and
  Timothy~E. McAlindon.
\newblock Association of knee injuries with accelerated knee osteoarthritis
  progression: Data from the osteoarthritis initiative.
\newblock {\em Arthritis Care {\&} Research}, 66(11):1673--1679, October 2014.

\bibitem{dunn2004t2}
Timothy~C Dunn, Ying Lu, Hua Jin, Michael~D Ries, and Sharmila Majumdar.
\newblock T2 relaxation time of cartilage at mr imaging: comparison with
  severity of knee osteoarthritis.
\newblock {\em Radiology}, 232(2):592--598, 2004.

\bibitem{Eijgenraam2019}
Susanne~M. Eijgenraam, Akshay~S. Chaudhari, Max Reijman, Sita M.~A.
  Bierma-Zeinstra, Brian~A. Hargreaves, Jos Runhaar, Frank W.~J. Heijboer,
  Garry~E. Gold, and Edwin H.~G. Oei.
\newblock Time-saving opportunities in knee osteoarthritis: T2 mapping and
  structural imaging of the knee using a single 5-min {MRI} scan.
\newblock {\em European Radiology}, 30(4):2231--2240, December 2019.

\bibitem{epperson2013creation}
K~Epperson.
\newblock Creation of fully sampled mr data repository for compressed sensing
  of the knee.
\newblock In {\em SMRT Annual Meeting}, volume~22, page~1, 2013.

\bibitem{Hammernik2017}
Kerstin Hammernik, Teresa Klatzer, Erich Kobler, Michael~P. Recht, Daniel~K.
  Sodickson, Thomas Pock, and Florian Knoll.
\newblock Learning a variational network for reconstruction of accelerated
  {MRI} data.
\newblock {\em Magnetic Resonance in Medicine}, 79(6):3055--3071, November
  2017.

\bibitem{heimann2010segmentation}
Tobias Heimann, Bryan~J Morrison, Martin~A Styner, Marc Niethammer, and
  S~Warfield.
\newblock Segmentation of knee images: a grand challenge.
\newblock In {\em Proc. MICCAI Workshop on Medical Image Analysis for the
  Clinic}, pages 207--214. Beijing, China, 2010.

\bibitem{Hunter2011}
D.J. Hunter, A.~Guermazi, G.H. Lo, A.J. Grainger, P.G. Conaghan, R.M. Boudreau,
  and F.W. Roemer.
\newblock Evolution of semi-quantitative whole joint assessment of knee {OA}:
  {MOAKS} ({MRI} osteoarthritis knee score).
\newblock {\em Osteoarthritis and Cartilage}, 19(8):990--1002, August 2011.

\bibitem{Knoll2020}
Florian Knoll, Tullie Murrell, Anuroop Sriram, Nafissa Yakubova, Jure Zbontar,
  Michael Rabbat, Aaron Defazio, Matthew~J. Muckley, Daniel~K. Sodickson,
  C.~Lawrence Zitnick, and Michael~P. Recht.
\newblock Advancing machine learning for {MR} image reconstruction with an open
  competition: Overview of the 2019 {fastMRI} challenge.
\newblock {\em Magnetic Resonance in Medicine}, 84(6):3054--3070, June 2020.

\bibitem{Knoll2020dataset}
Florian Knoll, Jure Zbontar, Anuroop Sriram, Matthew~J. Muckley, Mary Bruno,
  Aaron Defazio, Marc Parente, Krzysztof~J. Geras, Joe Katsnelson, Hersh
  Chandarana, Zizhao Zhang, Michal Drozdzalv, Adriana Romero, Michael Rabbat,
  Pascal Vincent, James Pinkerton, Duo Wang, Nafissa Yakubova, Erich Owens,
  C.~Lawrence Zitnick, Michael~P. Recht, Daniel~K. Sodickson, and Yvonne~W.
  Lui.
\newblock {fastMRI}: A publicly available raw k-space and {DICOM} dataset of
  knee images for accelerated {MR} image reconstruction using machine learning.
\newblock {\em Radiology: Artificial Intelligence}, 2(1):e190007, January 2020.

\bibitem{Kogan2017}
Feliks Kogan, Evan Levine, Akshay~S. Chaudhari, Uchechukwuka~D. Monu, Kevin
  Epperson, Edwin~H.G. Oei, Garry~E. Gold, and Brian~A. Hargreaves.
\newblock Simultaneous bilateral-knee {MR} imaging.
\newblock {\em Magnetic Resonance in Medicine}, 80(2):529--537, December 2017.

\bibitem{Menze2015}
Bjoern~H. Menze, Andras Jakab, Stefan Bauer, Jayashree Kalpathy-Cramer, Keyvan
  Farahani, Justin Kirby, Yuliya Burren, Nicole Porz, Johannes Slotboom, Roland
  Wiest, Levente Lanczi, Elizabeth Gerstner, Marc-Andre Weber, Tal Arbel,
  Brian~B. Avants, Nicholas Ayache, Patricia Buendia, D.~Louis Collins, Nicolas
  Cordier, Jason~J. Corso, Antonio Criminisi, Tilak Das, Herve Delingette,
  Cagatay Demiralp, Christopher~R. Durst, Michel Dojat, Senan Doyle, Joana
  Festa, Florence Forbes, Ezequiel Geremia, Ben Glocker, Polina Golland,
  Xiaotao Guo, Andac Hamamci, Khan~M. Iftekharuddin, Raj Jena, Nigel~M. John,
  Ender Konukoglu, Danial Lashkari, Jose~Antonio Mariz, Raphael Meier, Sergio
  Pereira, Doina Precup, Stephen~J. Price, Tammy~Riklin Raviv, Syed M.~S. Reza,
  Michael Ryan, Duygu Sarikaya, Lawrence Schwartz, Hoo-Chang Shin, Jamie
  Shotton, Carlos~A. Silva, Nuno Sousa, Nagesh~K. Subbanna, Gabor Szekely,
  Thomas~J. Taylor, Owen~M. Thomas, Nicholas~J. Tustison, Gozde Unal, Flor
  Vasseur, Max Wintermark, Dong~Hye Ye, Liang Zhao, Binsheng Zhao, Darko Zikic,
  Marcel Prastawa, Mauricio Reyes, and Koen~Van Leemput.
\newblock The multimodal brain tumor image segmentation benchmark ({BRATS}).
\newblock {\em {IEEE} Transactions on Medical Imaging}, 34(10):1993--2024,
  October 2015.

\bibitem{mildenberger2002introduction}
Peter Mildenberger, Marco Eichelberg, and Eric Martin.
\newblock Introduction to the dicom standard.
\newblock {\em European radiology}, 12(4):920--927, 2002.

\bibitem{Miller2016}
Karla~L Miller, Fidel Alfaro-Almagro, Neal~K Bangerter, David~L Thomas, Essa
  Yacoub, Junqian Xu, Andreas~J Bartsch, Saad Jbabdi, Stamatios~N Sotiropoulos,
  Jesper L~R Andersson, Ludovica Griffanti, Gwenaëlle Douaud, Thomas~W Okell,
  Peter Weale, Iulius Dragonu, Steve Garratt, Sarah Hudson, Rory Collins, Mark
  Jenkinson, Paul~M Matthews, and Stephen~M Smith.
\newblock Multimodal population brain imaging in the {UK} biobank prospective
  epidemiological study.
\newblock {\em Nature Neuroscience}, 19(11):1523--1536, September 2016.

\bibitem{milletari2016v}
Fausto Milletari, Nassir Navab, and Seyed-Ahmad Ahmadi.
\newblock V-net: Fully convolutional neural networks for volumetric medical
  image segmentation.
\newblock In {\em 2016 fourth international conference on 3D vision (3DV)},
  pages 565--571. IEEE, 2016.

\bibitem{miyazaki2002dynamic}
T~Miyazaki, M~Wada, H~Kawahara, M~Sato, H~Baba, and S~Shimada.
\newblock Dynamic load at baseline can predict radiographic disease progression
  in medial compartment knee osteoarthritis.
\newblock {\em Annals of the rheumatic diseases}, 61(7):617--622, 2002.

\bibitem{MONAI_Consortium_MONAI_Medical_Open_2020}
{MONAI Consortium}.
\newblock {MONAI: Medical Open Network for AI}, 3 2020.

\bibitem{monu2017cluster}
Uchechukwuka~D Monu, Caroline~D Jordan, Bonnie~L Samuelson, Brian~A Hargreaves,
  Garry~E Gold, and Emily~J McWalter.
\newblock Cluster analysis of quantitative mri t2 and t1$\rho$ relaxation times
  of cartilage identifies differences between healthy and acl-injured
  individuals at 3t.
\newblock {\em Osteoarthritis and cartilage}, 25(4):513--520, 2017.

\bibitem{Muckley2021}
Matthew~J. Muckley, Bruno Riemenschneider, Alireza Radmanesh, Sunwoo Kim, Geunu
  Jeong, Jingyu Ko, Yohan Jun, Hyungseob Shin, Dosik Hwang, Mahmoud Mostapha,
  Simon Arberet, Dominik Nickel, Zaccharie Ramzi, Philippe Ciuciu, Jean-Luc
  Starck, Jonas Teuwen, Dimitrios Karkalousos, Chaoping Zhang, Anuroop Sriram,
  Zhengnan Huang, Nafissa Yakubova, Yvonne~W. Lui, and Florian Knoll.
\newblock Results of the 2020 {fastMRI} challenge for machine learning {MR}
  image reconstruction.
\newblock {\em {IEEE} Transactions on Medical Imaging}, pages 1--1, 2021.

\bibitem{mundermann2004potential}
Anne M{\"u}ndermann, Chris~O Dyrby, Debra~E Hurwitz, Leena Sharma, and Thomas~P
  Andriacchi.
\newblock Potential strategies to reduce medial compartment loading in patients
  with knee osteoarthritis of varying severity: reduced walking speed.
\newblock {\em Arthritis \& Rheumatism: Official Journal of the American
  College of Rheumatology}, 50(4):1172--1178, 2004.

\bibitem{ong2018mridata}
F~Ong, S~Amin, S~Vasanawala, and M~Lustig.
\newblock Mridata. org: An open archive for sharing mri raw data.
\newblock In {\em Proc. Intl. Soc. Mag. Reson. Med}, volume~26, page~1, 2018.

\bibitem{ong2019sigpy}
Frank Ong and Michael Lustig.
\newblock Sigpy: a python package for high performance iterative
  reconstruction.
\newblock {\em Proceedings of the International Society of Magnetic Resonance
  in Medicine, Montr{\'e}al, QC}, 4819, 2019.

\bibitem{Pace2015}
Danielle~F. Pace, Adrian~V. Dalca, Tal Geva, Andrew~J. Powell, Mehdi~H.
  Moghari, and Polina Golland.
\newblock Interactive whole-heart segmentation in congenital heart disease.
\newblock In {\em Lecture Notes in Computer Science}, pages 80--88. Springer
  International Publishing, 2015.

\bibitem{parisot1995dicom}
Charles Parisot.
\newblock The dicom standard.
\newblock {\em The International Journal of Cardiac Imaging}, 11(3):171--177,
  1995.

\bibitem{Peterfy2008}
C.G. Peterfy, E.~Schneider, and M.~Nevitt.
\newblock The osteoarthritis initiative: report on the design rationale for the
  magnetic resonance imaging protocol for the knee.
\newblock {\em Osteoarthritis and Cartilage}, 16(12):1433--1441, December 2008.

\bibitem{Petersen2009}
R.~C. Petersen, P.~S. Aisen, L.~A. Beckett, M.~C. Donohue, A.~C. Gamst, D.~J.
  Harvey, C.~R. Jack, W.~J. Jagust, L.~M. Shaw, A.~W. Toga, J.~Q. Trojanowski,
  and M.~W. Weiner.
\newblock Alzheimer{\textquotesingle}s disease neuroimaging initiative
  ({ADNI}): Clinical characterization.
\newblock {\em Neurology}, 74(3):201--209, December 2009.

\bibitem{pruessmann1999sense}
Klaas~P Pruessmann, Markus Weiger, Markus~B Scheidegger, and Peter Boesiger.
\newblock Sense: sensitivity encoding for fast mri.
\newblock {\em Magnetic Resonance in Medicine: An Official Journal of the
  International Society for Magnetic Resonance in Medicine}, 42(5):952--962,
  1999.

\bibitem{ronneberger2015u}
Olaf Ronneberger, Philipp Fischer, and Thomas Brox.
\newblock U-net: Convolutional networks for biomedical image segmentation.
\newblock In {\em International Conference on Medical image computing and
  computer-assisted intervention}, pages 234--241. Springer, 2015.

\bibitem{sandino2020compressed}
Christopher~M Sandino, Joseph~Y Cheng, Feiyu Chen, Morteza Mardani, John~M
  Pauly, and Shreyas~S Vasanawala.
\newblock Compressed sensing: From research to clinical practice with deep
  neural networks: Shortening scan times for magnetic resonance imaging.
\newblock {\em IEEE signal processing magazine}, 37(1):117--127, 2020.

\bibitem{sriram2020end}
Anuroop Sriram, Jure Zbontar, Tullie Murrell, Aaron Defazio, C~Lawrence
  Zitnick, Nafissa Yakubova, Florian Knoll, and Patricia Johnson.
\newblock End-to-end variational networks for accelerated mri reconstruction.
\newblock In {\em International Conference on Medical Image Computing and
  Computer-Assisted Intervention}, pages 64--73. Springer, 2020.

\bibitem{Sveinsson2017}
B.~Sveinsson, A.S. Chaudhari, G.E. Gold, and B.A. Hargreaves.
\newblock A simple analytic method for estimating t2 in the knee from {DESS}.
\newblock {\em Magnetic Resonance Imaging}, 38:63--70, May 2017.

\bibitem{vanBeek2018}
Edwin~J.R. van Beek, Christiane Kuhl, Yoshimi Anzai, Patricia Desmond,
  Richard~L. Ehman, Qiyong Gong, Garry Gold, Vikas Gulani, Margaret
  Hall-Craggs, Tim Leiner, C.C.~Tschoyoson Lim, James~G. Pipe, Scott Reeder,
  Caroline Reinhold, Marion Smits, Daniel~K. Sodickson, Clare Tempany,
  H.~Alberto Vargas, and Meiyun Wang.
\newblock Value of {MRI} in medicine: More than just another test?
\newblock {\em Journal of Magnetic Resonance Imaging}, 49(7):e14--e25, August
  2018.

\bibitem{ying2007joint}
Leslie Ying and Jinhua Sheng.
\newblock Joint image reconstruction and sensitivity estimation in sense
  (jsense).
\newblock {\em Magnetic Resonance in Medicine: An Official Journal of the
  International Society for Magnetic Resonance in Medicine}, 57(6):1196--1202,
  2007.

\bibitem{zhao2021fastmri+}
Ruiyang Zhao, Burhaneddin Yaman, Yuxin Zhang, Russell Stewart, Austin Dixon,
  Florian Knoll, Zhengnan Huang, Yvonne~W Lui, Michael~S Hansen, and Matthew~P
  Lungren.
\newblock fastmri+: Clinical pathology annotations for knee and brain fully
  sampled multi-coil mri data.
\newblock {\em arXiv preprint arXiv:2109.03812}, 2021.

\end{thebibliography}
}

\clearpage

\appendix

\begin{table}
    \centering
    \caption{qDESS scan acquisition parameters for the SKM-TEA dataset. RO -- readout, PE -- phase encode, TE -- echo time, TR -- repetition time.}
    \label{tbl:scan-parameters}
    \begin{tabular}{l|c}
    \toprule
         Matrix (RO $\times$ PE) & 416$\times$512 \\
         Resolution (mm\textsuperscript{2}) & 0.38$\times$0.31\\
         TE - Echo 1 (ms) & 5.7 \\
         TE - Echo 2 (ms) & 30.1 \\
         Number of Echoes & 2 \\
         TR (ms) & 17.9 \\
         Flip Angle (\textdegree) &  20\\
         Parallel Imaging & 2$\times$1 \\
    \bottomrule
    \end{tabular}
\end{table}

\section{Dataset: Additional Details}
\label{app:dataset}

\subsection{Acquisition Parameters}
\label{app:dataset-acq-params}
All shared scan parameters are shown in Table \ref{tbl:scan-parameters}. qDESS scans were acquired with 2x1 parallel imaging with multiple receiver coils. Scans were acquired with 15 or 16 coils. Number of slices were varied based on the knee size, ranging from 80 to 88 slices.

\subsection{ZIP2 Zero-Padding}
\label{app:dataset-zero-padding}
K-space data for each scan was zero-padded along the readout dimension to $k_y \times k_z$ matrix size of 512$\times$512. The data was also zero-padded along the slice dimension so as to double the matrix size along this dimension, following the ZIP2 convention. The resulting  $k_x \times k_y \times k_z$ matrix size is 512$\times$512$\times$($2s$), where $s$ is the number of slices acquired.

The raw data distributed publicly is zero-padded. When undersampling this data, only the true acquisition region (416$\times$512) should be undersampled to the extent corresponding to the prescribed acceleration. The undersampling masks are generated such that the undersampling occurs only among the data acquisition region -- i.e. they do not include zero-padded region.

\subsection{Gradient-Warping Correction for Segmentations}
\label{app:grad-warp}
Scanner-generated DICOM images undergo vendor- and scanner-specific gradient warping to correct for gradient imperfections (between the nominal and actual magnetic fields) that results in a non-linear spatial deformation of the reconstructed image. As a result, segmentations annotated on the gradient-warped DICOM images did not overlap with appropriate regions in the SENSE-reconstructed images. To correct for this, a regional b-spline registration algorithm (available in DOSMA \cite{desai2019dosma}) was used to register the DICOM images and the corresponding segmentation masks to SENSE reconstructions for each scan. We refer to these as the \textit{gradient-warp-corrected segmentations}. All analysis or end-to-end inter-operation between SENSE reconstructions and tissue segmentations should use the gradient-warp-corrected segmentation. From manual inspection, no such process was required for the coarser bounding box pathology labels.

\RV{
\subsection{Annotator Details}
\label{app:annotator-details}
Detection bounding boxes and tissue segmentations were created by two researchers with 3-4 years experience with knee MR image interpretation, supervised by two board-certified musculoskeletal radiologists with 26 and 24 years of experience. All four individuals had semi-structured clinical radiology reports which were used to instruct bounding box labels. Each scan was annotated by a single annotator, such that no one scan had labels from multiple annotators.

For ground-truth cartilage and meniscus segmentations, annotators used both qDESS echoes that provide separate image contrasts to distinguish between the neighboring cartilage and meniscus pixels, as well as additional tissues such as bone, muscle, and joint fluid. Segmentations were performed slice-by-slice in the sagittal plane and volumetric consistency was enforced by correcting segmentations in the axial and coronal planes in the ITK-SNAP software. Every image volume segmentation was quality controlled by the two researchers with 4 and 3 years experience with knee MR image interpretation.
}

\subsection{Distribution, Hosting, and Maintenance}
\label{app:dataset-distribution}

All public data is distributed under the Stanford University School of Medicine (\url{http://www.stanford.edu/site/terms/}) license and the terms listed for the Lower Extremity Radiographs dataset (\url{https://aimi.stanford.edu/lera-lower-extremity-radiographs-2}). Data is hosted and maintained by the authors and Microsoft Azure as part of a partnership with the Stanford Center for Artificial Intelligence in Medicine and Imaging. All data and corresponding artifacts (annotations, etc.) will be semantically versioned and available for future use.

Instructions for downloading and using the dataset, versioned data splits and annotations, starter code, and baselines can be found on the dataset GitHub page: \url{https://github.com/StanfordMIMI/skm-tea}.

\subsection{Usage}
\label{app:dataset-usage}
Table \ref{tbl:data-per-track} summarizes the available data and the tracks with which they are compatible.

\textbf{Raw Data Track}: All raw data and artifacts originating from this data (sensitivity maps, SENSE reconstructions, etc.) should be used solely in the Raw Data Track. For segmentation-related analysis in this track, gradient-warp-corrected segmentations should be used in place of DICOM segmentations. For reconstruction, all evaluation results should be reported on data undersampled using the precomputed undersampling masks (at the appropriate acceleration) that are distributed with the raw data. Complex-valued SENSE reconstructed images should be used as ground-truth images for reconstruction evaluation.

\textbf{DICOM Track}: This track pertains to all tasks enabled by DICOM images and their artifacts (e.g. DICOM segmentations). For segmentation-related analysis, DICOM segmentations should be used. DICOM images should not be used for any part of the reconstruction task.

\begin{table}
    \centering
  \caption{Available data, whether it is a model input or output, the tracks to be used with, \RV{and corresponding tasks. N/A indicates the data should not be used for benchmarking any task.}}
  \label{tbl:data-per-track}
  \begin{tabular}{lcccc}
    \toprule
    Data     &  Data Type  &   Raw Data Track     & DICOM Track & \RV{Task} \\
    \midrule
    Raw data (k-space)                     &  Input   &    \cmark & \xmark & \RV{Recon}\\
    Undersampling masks (per-scan)         &  Input   &    \cmark & \xmark & \RV{Recon} \\
    Sensitivity maps                       &  Input   &    \cmark & \xmark & \RV{Recon} \\
    SENSE reconstruction                   &  Output  &    \cmark & \xmark & \RV{Recon, Seg, Detection} \\
    DICOM images                           &  Input   &    \xmark & \cmark & \RV{Seg, Detection} \\
    DICOM $T_2$ parametric maps            &  Input   &    \xmark & \xmark & \RV{N/A} \\
    DICOM segmentations                    &  Output  &    \xmark & \cmark & \RV{Seg} \\
    Gradient-warp-corrected segmentations  &  Output  &    \cmark & \xmark & \RV{Seg} \\
    Pathology bounding boxes               &  Output  &    \cmark & \cmark & \RV{Detection} \\
    \bottomrule
  \end{tabular}
\end{table}

\subsection{Author Statement}
We, the authors, confirm that we bear all responsibility in case of violation of rights, etc. Public data are distributed under the Stanford University School of Medicine (\url{http://www.stanford.edu/site/terms/}) license and the terms listed for the Lower Extremity Radiographs dataset (\url{https://aimi.stanford.edu/lera-lower-extremity-radiographs-2}).

\section{Tissue Subregions}
\label{app:tissue-subregions}
In this appendix, we detail the relevance of subregional tissue analysis in qMRI and the method by which different tissue subregions are extracted.

\subsection{Relevant Tissue Subregions}
\label{app:tissue-subregions-relevant}
Acute knee injuries and knee degeneration are predominantly localized processes, where specific subregions of the knee undergo more change than others \cite{Driban2014}. To quantify local qMRI parameter profiles, specific subregions of relevant tissues must be precisely segmented. Recent work has shown that subregions in articular cartilage and the meniscus are sensitive to early-onset of degenerative diseases, such as osteoarthritis \cite{dunn2004t2}. Thus, we include subregional analysis of the segmented tissues in our proposed qMRI evaluation framework, as was previously described \cite{monu2017cluster}.

\subsection{Subregion Extraction}
\label{app:tissue-subregions-extraction}

Segmentations for each of the four nominal tissues (patellar cartilage, femoral cartilage, tibial cartilage, and meniscus) served as the base ROIs. These segmentations were then divided into subregions using shape-based priors and center-of-mass (COM) estimates, which are detailed below. 
% Sub-regions were abbreviated based on planar compartments along the principal axes in the following order - axial, coronal, and sagittal. For example, a subregion from axial compartment \textit{a}, coronal compartment \textit{R}, and sagittal compartment \textit{S} would have the abbreviation \textit{aR-S}.
Subregions were abbreviated based on their anatomical location. For example, a subregion for the deep cartilage compartment \textit{d}, anterior part of the knee \textit{A}, and the lateral condyle \textit{L} would have the abbreviation \textit{dA-L}. All sub-regions were extracted automatically using DOSMA (v0.1.0).

\textbf{Patellar cartilage:} Patellar cartilage was divided into four subregions: deep-lateral (d-L), deep-medial (d-M), superficial-lateral (s-L), and superficial-medial (s-M). The medial/lateral boundary was determined by the COM of the patellar cartilage segmentation along the sagittal plane. The deep/superficial boundary was computed by finding the midpoint of each column in the patellar cartilage segmentation along the coronal plane.

\textbf{Femoral cartilage:} Femoral cartilage segmentations were divided into a total of 12 regions along the three primary axes: deep/superficial (d/s), anterior/central/posterior(A/C/P), and medial/lateral (M/L). Resulting subregions were named following the nomenclature of these axes; for example, \textit{dA-M} corresponds to the deep anterior cartilage in the medial compartment. A/C/P and M/L compartments were delineated based on COM measurements of the base segmentation. The deep-superficial boundary was computed using the "unrolling technique" \cite{monu2017cluster}, where the boundary is determined by the midpoint between radii of concentric cylindrical fits to the femoral cartilage shape.

\textbf{Tibial cartilage:} Tibial cartilage was also divided into 12 subregions: inferior/superior (i/sup), A/C/P, and M/L. A/C/P and M/L compartments were divided based on the COM of the base segmentation. The inferior/superior (i/sup) boundary was determined by finding the COM for each tibial cartilage column in the axial direction \cite{crowder2021characterizing}.

\textbf{Meniscus}: The meniscus was divided into medial/lateral compartments using COM between the two compartments.

% Include extra information in the appendix. This section will often be part of the supplemental material. Please see the call on the NeurIPS website for links to additional guides on dataset publication.

\RV{
\section{qMRI $T_2$ Pipeline}
\label{app:qMRI-evaluation-pipeline}
In this section, we discuss the recommended pipeline for computing ground truth and predicted $T_2$ estimates for benchmarks in the Raw Data Track and DICOM Track. For fairness of comparison, this pipeline should be used when comparing results from future benchmarks and methods for the SKM-TEA dataset to results detailed in this work.

\textbf{Raw Data Track -- Reconstruction:} Image reconstruction methods were used to generate reconstructions for echo 1 (E1) and echo 2 (E2). Two $T_2$ parameter maps were computed for each scan, one using the network reconstruction and the other using the ground-truth SENSE reconstruction. Ground-truth gradient-warp-corrected segmentations were used to identify relevant tissues in both parameter maps. Differences in $T_2$ estimates were computed between regional $T_2$ estimates from the two parameter maps.

\textbf{DICOM Track -- Segmentation:} For each scan, a single $T_2$ parameter map was computed from the DICOM images. Image segmentatation methods were used to generate predicted masks for relevant tissues. Differences in $T_2$ estimates were computed between regional $T_2$ estimates extracted using the ground truth mask and the predicted mask.
}

\section{Training Details}
\label{app:training-details}

In this section, we cover details pertaining to the model architecture, training setup, and compute resources used for the Raw Data Track reconstruction and DICOM Track segmentation benchmarks. Model training and evaluation was conducted in PyTorch. Code to reproduce all results with detailed instructions and evolving benchmarks are available at \url{https://github.com/StanfordMIMI/skm-tea}.

\subsection{Raw Data Track -- Reconstruction}

\textbf{Training setup:} In this problem, models are trained to reconstruct complex-valued, 2D undersampled axial ($k_y \times k_z$) slices for both qDESS echoes. Scans are undersampled using 2D Poisson Disc undersampling at acceleration factors of $R$=6x, 8x. Models are trained separately at each acceleration and evaluated on simulated undersampled scans at the respective acceleration. All models were trained with the complex-L1 loss with a fixed random seed.

\textbf{Data normalization}: Input data is normalized by dividing by the 95\textsuperscript{th} percentile of the magnitude image. Outputs are re-normalized by undoing the scaling operation prior to computation of evaluation metrics. Outputs are not re-normalized during training, prior to computing the training loss.

\textbf{Undersampling masks:} During training, 100,000 undersampling masks are precomputed and cached to ensure all training runs use the same set of undersampling masks. For evaluation, each scan in the test dataset is prescribed a fixed undersampling mask (for the specific acceleration) that is distributed as part of the dataset. All masks are generated such that only the acquisition region is undersampled - i.e. all zero-padded regions in the kspace are not included in the generated undersampling mask.

\textbf{U-Net baseline:} We consider a 2D U-Net model, a popular model for fully convolutional and image-to-image tasks, following the implementation in \cite{Knoll2020} as one baseline architecture. This U-Net implementation has four max pooling layers with compounding number of channels (32, 64, 128, 256, 512), instance normalization, and leaky-relu activation with slope $\alpha$=-0.2. All U-Net models were trained for 20 epochs using the Adam optimizer with the following hyperparameters: batch size 24, learning rate $\eta$=1e-3, weight decay 1e-4. 

\textbf{Unrolled baseline}: We consider the 2D proximal-gradient unrolled network, which has achieved state-of-the-art performance on MRI reconstruction tasks, as another baseline architecture. We follow the unrolled network in \cite{sandino2020compressed} with minimal hyperparameter changes. Each unrolled block consists of a shallow residual network with two, 128-channel residual blocks with relu activation. The network consists of a total of eight sequential unrolled blocks with weighted data consistency between each block. All unrolled models were trained for 20 epochs using the Adam optimizer. Due to hardware memory constraints, the same batch size as the U-Net could not be used. Instead a smaller batch size of 4 with 6 gradient accumulation steps was used so that the effective batch size is the same as that of U-Net baselines. A learning rate of $\eta$=8e-4 and weight decay of 1e-4 were used.

\textbf{Hardware}: All models were trained on Titan RTX (24GB) or GCP-supported Titan V100 (16GB) GPUs. Models trained on Titan RTX GPUs were constrained so that the total available memory was identical to the Titan V100 GPU (16GB).

\subsection{DICOM Track -- Segmentation}

\textbf{Training setup:} In this problem, models are trained segment patellar cartilage, femoral cartilage, tibial cartilage, and the meniscus from 2D sagittal slices of the DICOM images. All models were trained with a soft Dice loss with a fixed random seed.

\textbf{Input normalization:} All inputs are zero-mean, unit standard deviation normalized by mean and standard deviation values computed over the full volume of the echo. For multi-channel inputs (i.e. E1$\oplus$E2), each channel is normalized independently. Root-sum-of-squares (RSS) inputs are normalized by mean and standard deviation values computed on the RSS volume.

\textbf{V-Net baseline:} Another baseline used a 2D V-Net architecture as implemented in MONAI \cite{MONAI_Consortium_MONAI_Medical_Open_2020}. This network has 4 pooling layers with doubling number of channels after each pooling step (16, 32, 64, 128, 256). Neither dropout nor early stopping was not used.

\textbf{U-Net baseline:} One baseline used a 2D U-Net architecture as implemented in \cite{desai2019technical}. This network has 5 pooling layers with doubling number of channels after each pooling step (32, 64, 128, 256, 512, 1024). Convolutional blocks at each encoder and decoder level are composed of two convolutional layers each with relu activations followed by a batch normalization layer.

\textbf{Default training hyperparameters:} Models were trained using the Adam optimizer with initial learning rate $\eta_0$=1e-3, minimum learning rate $\eta_{min}$=1e-8, and step decay by 0.9x every 2 epochs. A maximum training length of 100 epochs was prescribed with early stopping ($\delta$=1e-5, $\tau$=12 epochs). Training batch size was set to 16 without gradient accumulation. All benchmarks used these hyperparameters unless otherwise mentioned.

\textbf{Hardware:} All models were trained on Quadro RTX 8000 (48GB) GPUs, but were constrained to only use 24GB memory.

\section{Additional Results}
\label{app:additional-results}

\begin{table}
  \caption{Performance of U-Net segmentation models measured by standard ML pixel and surface segmentation metrics with absolute $T_2$ error. Models were trained with echo 1 only (E1), echo 2 only (E2), multi-channel echo1 and echo2 (E1$\oplus$E2), and the root-sum-of-squares (RSS) of both echoes.}
  \label{tbl:unet-perf}
  \centering
\begin{tabular}{llcccc}
\toprule
     Metric & Tissue &    E1 &    E2 & E1$\oplus$E2 & RSS \\
\midrule
\multirow{4}{*}{DSC} & Patellar Cartilage &  0.87 (0.097) &   0.85 (0.11) &         0.87 (0.088) &   0.87 (0.10) \\
                  & Femoral Cartilage &  0.88 (0.035) &  0.86 (0.032) &         0.88 (0.033) &  0.88 (0.032) \\
                  & Tibial Cartilage &  0.86 (0.036) &  0.82 (0.049) &         0.86 (0.041) &  0.86 (0.037) \\
                  & Meniscus &  0.84 (0.062) &  0.82 (0.047) &         0.84 (0.067) &  0.84 (0.065) \\
\midrule
\multirow{4}{*}{ASSD (mm)} & Patellar Cartilage &    0.86 (1.4) &   0.58 (0.82) &           0.84 (1.6) &    1.52 (2.0) \\
                  & Femoral Cartilage &   0.36 (0.25) &   0.36 (0.20) &          0.40 (0.60) &   0.33 (0.19) \\
                  & Tibial Cartilage &   0.46 (0.29) &   0.45 (0.18) &           0.66 (1.2) &   0.66 (0.70) \\
                  & Meniscus &   0.63 (0.42) &   0.64 (0.29) &           0.91 (1.5) &    1.24 (1.4) \\
\midrule
\multirow{4}{*}{Abs T2 Error (ms)} & Patellar Cartilage &   0.70 (0.59) &   0.92 (0.95) &          0.71 (0.63) &   0.75 (0.59) \\
                  & Femoral Cartilage &   0.50 (0.36) &   0.92 (0.50) &          0.53 (0.37) &   0.51 (0.36) \\
                  & Tibial Cartilage &   0.49 (0.47) &   0.98 (0.66) &          0.51 (0.53) &   0.50 (0.52) \\
                  & Meniscus &   0.60 (0.78) &    1.07 (1.0) &           0.96 (1.0) &   0.65 (0.74) \\
\bottomrule
\end{tabular}
\end{table}

\begin{table}[]
    \centering
    \caption{Performance [mean (standard deviation)] of segmentation models on the DICOM Track as measured by volumetric overlap error (VOE), coefficient-of-variation (CV), and $T_2$ error (in milliseconds). $T_2$ error values are not bolded as it is unclear if better performance is characterized by smaller bias or lower variance.}
    \label{tbl:seg-metrics-extra}
\begin{tabular}{llllll}
\toprule
\toprule
  Metric & Tissue &      V-Net (E1) &      V-Net (E2) & V-Net (E1$\oplus$E2) &     V-Net (RSS) \\
\midrule
\multirow{4}{*}{VOE} & Patellar Cartilage &   0.205 (0.108) &   0.242 (0.127) &        0.193 (0.102) &   0.201 (0.114) \\
              & Femoral Cartilage &  0.214 (0.0551) &  0.237 (0.0496) &       0.205 (0.0457) &  0.210 (0.0511) \\
              & Tibial Cartilage &  0.241 (0.0547) &  0.288 (0.0675) &       0.238 (0.0516) &  0.238 (0.0515) \\
              & Meniscus &  0.259 (0.0818) &   0.289 (0.071) &       0.257 (0.0769) &  0.256 (0.0829) \\
\midrule
\multirow{4}{*}{CV} & Patellar Cartilage &  0.078 (0.0854) &  0.077 (0.0558) &       0.066 (0.0913) &  0.078 (0.0964) \\
              & Femoral Cartilage &  0.085 (0.0613) &   0.077 (0.059) &       0.076 (0.0539) &  0.080 (0.0573) \\
              & Tibial Cartilage &  0.095 (0.0691) &   0.092 (0.084) &       0.094 (0.0726) &  0.092 (0.0645) \\
              & Meniscus &  0.084 (0.0707) &  0.081 (0.0661) &       0.074 (0.0698) &  0.074 (0.0662) \\
\midrule
\multirow{4}{*}{T2 Error (ms)} & Patellar Cartilage &   0.486 (0.637) &   0.873 (0.928) &        0.531 (0.934) &   0.474 (0.673) \\
              & Femoral Cartilage &   0.172 (0.632) &   0.772 (0.657) &        0.287 (0.532) &   0.282 (0.624) \\
              & Tibial Cartilage &   0.215 (0.681) &   0.805 (0.831) &        0.208 (0.669) &   0.243 (0.698) \\
              & Meniscus &  -0.121 (0.813) &   -0.911 (1.01) &       -0.666 (0.857) &  -0.325 (0.821) \\
\midrule
\midrule
Metric & Tissue &      U-Net (E1) & U-Net (E1$\oplus$E2) &      U-Net (E2) &     U-Net (RSS) \\
\midrule

\multirow{4}{*}{VOE} & Patellar Cartilage &   0.219 (0.121) &        \textbf{0.216 (0.110)} &   0.245 (0.131) &   0.222 (0.118) \\
              & Femoral Cartilage &   0.220 (0.054) &        0.217 (0.052) &   0.246 (0.048) &   \textbf{0.216 (0.050)} \\
              & Tibial Cartilage &   \textbf{0.245 (0.053)} &        0.246 (0.060) &   0.297 (0.067) &   0.247 (0.055) \\
              & Meniscus &   \textbf{0.271 (0.084)} &        0.274 (0.089) &   0.300 (0.066) &   0.275 (0.088) \\
\midrule
\multirow{4}{*}{CV} & Patellar Cartilage &  0.0641 (0.085) &       0.0835 (0.101) &  \textbf{0.0640 (0.081)} &  0.0724 (0.083) \\
              & Femoral Cartilage &  0.0754 (0.060) &       0.0772 (0.057) &  \textbf{0.0746 (0.058)} &  0.0757 (0.057) \\
              & Tibial Cartilage &  0.0811 (0.058) &       0.0882 (0.076) &  \textbf{0.0799 (0.076)} &  0.0848 (0.064) \\
              & Meniscus &  \textbf{0.0716 (0.067)} &       0.0803 (0.082) &  0.0816 (0.063) &  0.0824 (0.076) \\

\midrule
\multirow{4}{*}{T2 Error (ms)} & Patellar Cartilage &   0.468 (0.790) &        0.510 (0.805) &   0.484 (1.23) &   0.441 (0.854) \\
              & Femoral Cartilage &   0.152 (0.596) &        0.388 (0.522) &   0.837 (0.639) &   0.237 (0.585) \\
              & Tibial Cartilage &   0.147 (0.666) &        0.179 (0.714) &   0.887 (0.783) &   0.180 (0.698) \\
              & Meniscus &  -0.360 (0.923) &       -0.954 (1.048) &   -1.06 (1.05) &  -0.499 (0.847) \\
\bottomrule
\end{tabular}
\end{table}

\begin{table}
\centering
  \caption{Performance [mean (standard deviation)] of qDESS reconstruction models with respect to $T_2$ estimates (in milliseconds) for articular cartilage and the meniscus localized with ground truth segmentations. Typical cartilage $T_2$ values are 30-40ms, while meniscus $T_2$ values are 10-15ms.}
  \label{tbl:recon-t2-error}
\begin{tabular}{llllll}
\toprule
  & Tissue & Patellar Cartilage & Femoral Cartilage & Tibial Cartilage &       Meniscus \\
Acc & Model &                    &                   &                  &                \\
\midrule
\multirow{6}{*}{6x} & U-Net (E1/E2) &       -1.93 (1.98) &     -0.228 (1.42) &     -1.17 (1.48) &   -2.70 (1.35) \\
  & U-Net (E1+E2) &      -0.231 (3.46) &       1.83 (2.51) &     0.201 (1.73) &   -1.88 (1.59) \\
  & U-Net (E1$\oplus$E2) &       -1.25 (1.96) &     -0.838 (1.08) &     -1.44 (1.16) &   -1.78 (1.03) \\
  & Unrolled (E1/E2) &     -0.516 (0.327) &    -0.765 (0.283) &    -1.03 (0.419) &  -2.48 (0.786) \\
  & Unrolled (E1+E2) &     -0.555 (0.269) &    -0.836 (0.319) &    -1.12 (0.444) &  -2.52 (0.780) \\
  & Unrolled (E1$\oplus$E2) &      -0.639 (2.09) &     -2.01 (0.917) &    -1.30 (0.650) &  -1.25 (0.910) \\
\midrule
\multirow{6}{*}{8x} & U-Net (E1/E2) &       -3.48 (1.74) &      -2.71 (1.38) &     -3.21 (1.24) &   -3.76 (1.10) \\
  & U-Net (E1+E2) &       0.335 (3.38) &       2.75 (2.42) &     0.153 (1.89) &   -2.24 (1.55) \\
  & U-Net (E1$\oplus$E2) &      -0.247 (1.68) &     -0.889 (1.29) &     -1.93 (1.39) &   -1.88 (2.45) \\
  & Unrolled (E1/E2) &     -0.702 (0.340) &    -0.866 (0.415) &    -1.20 (0.618) &  -2.78 (0.868) \\
  & Unrolled (E1+E2) &     -0.971 (0.419) &    -0.977 (0.421) &    -1.26 (0.590) &  -2.86 (0.882) \\
  & Unrolled (E1$\oplus$E2) &     -0.482 (0.449) &    -0.817 (0.717) &    -1.15 (0.931) &  -2.69 (0.998) \\
\bottomrule
\end{tabular}
\end{table}

\subsection{Additional Segmentation Baselines}
\label{app:additional-seg-baselines}
In addition to the V-Net models, we trained baseline segmentation models with the U-Net architecture. Dice, ASSD, and absolute $T_2$ error are reported in Table \ref{tbl:unet-perf}. Like V-Net, U-Net models trained on only the second echo (E2) performed considerably worse across all metrics. V-Net models achieved slightly higher performance among standard ML segmentation metrics for patellar cartilage, but had similar performance among $T_2$ error metrics.

Following the convention of previous segmentation challenges \cite{desai2021international, heimann2010segmentation, Menze2015}, we also compute volumetric overlap error (VOE) and coefficient of variation (CV). A summary of segmentation model performance on these metrics is shown in Table \ref{tbl:seg-metrics-extra}. Top performing models -- U-Net (E1), U-Net (E1 $\oplus$ E2), U-Net (RSS) -- achieved similar performance and outperformed U-Net (E2) across both metrics.

\subsection{$\mathbf{T_2}$ Error}
In addition to absolute $T_2$ error, we measure the standard $T_2$ error, which can help characterize the bias and variance of the errors in $T_2$ estimates.

\textbf{Raw Data Track - Reconstruction:} Table \ref{tbl:recon-t2-error} summarizes $T_2$ error for all benchmarked reconstruction models. All models except U-Net (E1+E2) underestimate $T_2$ across all tissues. While unrolled networks have lower variance in $T_2$ estimates than U-Net models, the bias of unrolled networks is often larger that that of U-Net models. Thus, unrolled models may be more precise in estimating $T_2$, but may still need to be optimized to reduce bias in these estimates.

\textbf{DICOM Track - Segmentation:} $T_2$ error profiles for different segmentation models are summarized in Table \ref{tbl:seg-metrics-extra}. Top performing models have low bias for femoral cartilage and tibial cartilage compared to the patellar cartilage and meniscus. Segmentations from all models overestimate $T_2$ for articular cartilage but underestimate $T_2$ for the meniscus. Variance in $T_2$ estimates is also the highest for patellar cartilage and the meniscus. Thus, $T_2$ estimates in both patellar cartilage and meniscus may be more sensitive to changes in segmentation quality than estimates in femoral cartilage or tibial cartilage.

\begin{table}[]
    \centering
    \caption{Concordance between standard reconstruction (pSNR, SSIM) and segmentation (DSC, ASSD) metrics versus absolute error in $T_2$ estimates across different tissue sub-regions as measured by absolute value of the Pearson's correlation coefficient ($|\rho|$). Tissue subregions are defined in Appendix \ref{app:tissue-subregions-extraction}. Values are averaged over all baseline reconstruction and segmentation models.}
    \label{tbl:t2-corr-subregion}
\begin{tabular}{llrrrr}
\toprule
         & Base &  ASSD (mm) &  DSC &  SSIM &  pSNR (dB) \\
MainTissue & Subregion &            &      &       &            \\
\midrule
Patellar Cartilage & d-L &       0.31 & 0.67 &  0.16 &       0.36 \\
         & d-M &       0.21 & 0.59 &  0.23 &       0.53 \\
         & s-L &       0.29 & 0.64 &  0.25 &       0.46 \\
         & s-M &       0.11 & 0.29 &  0.30 &       0.49 \\
\midrule
Femoral Cartilage & dA-L &       0.36 & 0.68 &  0.32 &       0.37 \\
         & dA-M &       0.15 & 0.29 &  0.27 &       0.36 \\
         & dC-L &       0.08 & 0.20 &  0.08 &       0.22 \\
         & dC-M &       0.10 & 0.21 &  0.10 &       0.30 \\
         & dP-L &       0.03 & 0.06 &  0.21 &       0.37 \\
         & dP-M &       0.02 & 0.05 &  0.25 &       0.31 \\
         & sA-L &       0.20 & 0.44 &  0.32 &       0.37 \\
         & sA-M &       0.12 & 0.29 &  0.34 &       0.42 \\
         & sC-L &       0.03 & 0.04 &  0.27 &       0.44 \\
         & sC-M &       0.03 & 0.08 &  0.42 &       0.54 \\
         & sP-L &       0.02 & 0.02 &  0.40 &       0.46 \\
         & sP-M &       0.10 & 0.09 &  0.35 &       0.31 \\
\midrule
Tibial Cartilage & iA-L &       0.14 & 0.31 &  0.21 &       0.35 \\
         & iA-M &       0.01 & 0.10 &  0.19 &       0.34 \\
         & iC-L &       0.10 & 0.22 &  0.23 &       0.32 \\
         & iC-M &       0.17 & 0.41 &  0.13 &       0.28 \\
         & iP-L &       0.00 & 0.19 &  0.25 &       0.38 \\
         & iP-M &       0.03 & 0.14 &  0.24 &       0.36 \\
         & supA-L &       0.11 & 0.35 &  0.31 &       0.40 \\
         & supA-M &       0.00 & 0.01 &  0.40 &       0.48 \\
         & supC-L &       0.12 & 0.36 &  0.41 &       0.47 \\
         & supC-M &       0.05 & 0.16 &  0.39 &       0.49 \\
         & supP-L &       0.03 & 0.18 &  0.35 &       0.41 \\
         & supP-M &       0.03 & 0.05 &  0.36 &       0.45 \\
\midrule
Meniscus & L &       0.08 & 0.17 &  0.20 &       0.18 \\
         & M &       0.05 & 0.25 &  0.17 &       0.13 \\
\bottomrule
\end{tabular}
\end{table}

\subsection{ML-$\mathbf{T_2}$ Metric Concordance in Tissue Sub-Regions}
\label{app:t2-corr-subregions}
As mentioned in \S\ref{app:tissue-subregions-relevant}, subregional qMRI analysis is a pivotal tool for understanding localized changes in tissue structure. To understand sensitivity of ML metrics to these biomarker-driven metrics, we quantify the concordance between standard ML metrics for image reconstruction and segmentation and subregional absolute $T_2$ error using Pearson's correlation coefficient. For reconstruction, the global pSNR and SSIM, which are standard metrics computed for image quality, were compared to subregional $T_2$ estimate errors. For segmentation, subregional $T_2$ estimate errors were compared to segmentation metrics computed on the corresponding parent tissue structure. For example, $T_2$ error in the deep-superficial-lateral compartment of femoral cartilage was compared to DSC and ASSD of femoral cartilage. Tissue subregions are computed using methods detailed in \S\ref{app:tissue-subregions-extraction}. Table \ref{tbl:t2-corr-subregion} summarizes the results.

SSIM and pSNR have very weak correlation with subregional $T_2$ error ($|\rho|\leq 0.42, 0.54$ respectively). Because these metrics measure the global image quality, they are likely not sensitive to local regional changes in image quality, and thus, may be even less sensitive to subregional $T_2$ error. ASSD is also very weakly correlated with $T_2$ error across all subregions ($|\rho|\leq0.36$). Because ASSD is a surface metric, it may not capture the changes in volumetric subregions of the image over which these estimates are computed. DSC is a volumetric metric, which may explain why the average correlation is stronger with DSC than with ASSD. However, despite its volumetric nature, DSC is weakly correlated with $T_2$ error among almost all subregions. In particular, it is weakly correlated along the subregions in the medial compartment, which are the most likely to undergo degeneration in chronic degenerative diseases such as osteoarthritis \cite{miyazaki2002dynamic,mundermann2004potential}.

This may further warrant the need for direct biomarker-based evaluation metrics that the qMRI evaluation framework enables.

\end{document}